\documentclass[11pt,epsf]{article}
 \usepackage{amsmath}
 \usepackage{graphicx}
 \usepackage[merge,numbers,compress]{natbib}
 \usepackage[T1]{fontenc}
 \usepackage{booktabs}
 \usepackage{xcolor} 
 \usepackage{xspace}
 \usepackage{dcolumn}
 \usepackage{placeins}
 \usepackage{amssymb}
 \usepackage{nicefrac}
 \usepackage[colorlinks=true,citecolor=blue!50!black,linkcolor=black]{hyperref}
 \usepackage{caption}
 \usepackage
 [subrefformat=parens,position=top,skip=-15pt,margin=15pt,justification=justified,singlelinecheck=false]
 {subcaption}
 \usepackage{array,multirow}

\makeatletter
\newcommand{\oset}[3][0ex]{%
  \mathrel{\mathop{#3}\limits^{
    \vbox to#1{\kern-3\ex@
    \hbox{\tiny$\scriptstyle#2$}\vss}}}}
\makeatother

\newcommand{\pb}{{\ensuremath\unskip\,\text{pb}}\xspace}

\def\be{\begin{equation}}
\def\ee{\end{equation}}

\newcommand{\Pj}{\ensuremath{\text{j}}\xspace}
\newcommand{\Pp}{\ensuremath{\text{p}}\xspace}

\newcommand{\Pc}{\ensuremath{\text{c}}\xspace}

\newcommand{\PW}{\ensuremath{\text{W}}\xspace}
\newcommand{\PZ}{\ensuremath{\text{Z}}\xspace}

\newcommand{\GeV}{\ensuremath{\,\text{GeV}}\xspace}
\newcommand{\TeV}{\ensuremath{\,\text{TeV}}\xspace}

\newcommand{\alphas}{\ensuremath{\alpha_\text{s}}\xspace}

\newcommand{\newc}{\newcommand}
\newc{\bi}{\begin{itemize}}
\newc{\ei}{\end{itemize}}
\newc{\benu}{\begin{enumerate}}
\newc{\eenu}{\end{enumerate}}
\newc{\bc}{\begin{center}}
\newc{\ec}{\end{center}}
\newc{\bfig}{\begin{figure}}
\newc{\efig}{\end{figure}}
\newc{\qbar}{\bar{q}}
\newc{\go}{\tilde{g}}
\newc{\PB}{\textsc{Powheg-Box}}

\newcolumntype{.}{D{.}{.}{-1}}
\newcolumntype{d}[1]{D{.}{.}{#1}}

\colorlet{tableoverheadcolor}{gray!37.5}
\colorlet{tableheadcolor}{gray!25}
\colorlet{tablerowcolor}{gray!12.5}

\newlength{\width}
\newlength{\height}

\def\draftdate{\relax}
\def\mda{\relax}
\def\mua{\relax}
\def\mla{\relax}
\def\draft{
\def\thtystars{******************************}
\def\sixtystars{\thtystars\thtystars}
\typeout{}
\typeout{\sixtystars**}
\typeout{* Draft mode!
         For final version remove \protect\draft\space in source file *}
\typeout{\sixtystars**}
\typeout{}
\def\draftdate{\today}
\def\mua{\marginpar[\boldmath\hfil$\uparrow$]%
                   {\boldmath$\uparrow$\hfil}\color{black}%
                    \typeout{marginpar: $\uparrow$}\ignorespaces}
\def\mda{\color{red}\marginpar[\boldmath\hfil$\downarrow$]%
                   {\boldmath$\downarrow$\hfil}%
                    \typeout{marginpar: $\downarrow$}\ignorespaces}
\def\mla{\marginpar[\boldmath\hfil$\rightarrow$]%
                   {\boldmath$\leftarrow $\hfil}%
                    \typeout{marginpar: $\leftrightarrow$}\ignorespaces}
\def\Mua{\marginpar[\boldmath\hfil$\Uparrow$]%
                   {\boldmath$\Uparrow$\hfil}\color{black}%
                    \typeout{marginpar: $\uparrow$}\ignorespaces}
\def\Mda{\color{red}\marginpar[\boldmath\hfil$\Downarrow$]%
                   {\boldmath$\Downarrow$\hfil}%
                    \typeout{marginpar: $\downarrow$}\ignorespaces}
\def\Mla{\marginpar[\boldmath\hfil\textcolor{red}{$\Rightarrow$}]%
                   {\boldmath\textcolor{red}{$\Leftarrow $}\hfil}%
                    \typeout{marginpar: $\leftrightarrow$}\ignorespaces}
\overfullrule 5pt
\oddsidemargin 15mm
\marginparwidth 29mm
}

\begin{document}

\title{\hfill ~\\[-30mm]
\phantom{h} \hfill\mbox{\small CAVENDISH--HEP--22/04, FR-PHENO-2022-04, Lund-22-24}
\\[1cm]
\vspace{13mm}   \textbf{Angular coefficients in $\PW+\Pj$ production \\
 at the LHC with high precision}}

\date{}
\author{
Mathieu Pellen$^{1\,}$\footnote{E-mail:  \texttt{mathieu.pellen@physik.uni-freiburg.de}},
Rene Poncelet$^{2\,}$\footnote{E-mail:  \texttt{poncelet@hep.phy.cam.ac.uk}},
Andrei Popescu$^{2\,}$\footnote{E-mail:  \texttt{popescu@hep.phy.cam.ac.uk}},
Timea Vitos$^{3\,}$\footnote{E-mail:  \texttt{timea.vitos@thep.lu.se}}
\\[9mm]
{\small\it $^1$ Albert-Ludwigs-Universit\"at Freiburg, Physikalisches Institut,} \\ %
{\small\it Hermann-Herder-Stra\ss e 3, D-79104 Freiburg, Germany}\\[3mm]
{\small\it $^2$ Cavendish Laboratory, University of Cambridge,} \\ %
{\small\it J.J. Thomson Avenue, Cambridge CB3 0HE, United Kingdom}\\[3mm]
{\small\it $^3$ Theoretical particle physics, Lund University,} \\ %
{\small\it Sölvegatan 14A, SE-223 62, Lund, Sweden}\\[3mm]
        }
\maketitle

\begin{abstract}
\noindent

The extraction of the W-boson mass, a fundamental parameter of the Standard Model, from hadron-hadron collision requires precise theory predictions.
In this regard, angular coefficients are crucial to model the dynamics of W-boson production.
In this work, we provide, for the first time, angular coefficients at NNLO QCD + NLO EW accuracy for finite transverse momentum W-boson at the LHC.
The corrections can reach up to $10\%$ in certain regions of phase space.
They are accompanied by a significant reduction of the scale uncertainty.
This work should, besides providing reference values for theory-data comparison, provide state-of-the-art theory input for $\PW$-boson mass measurements.

\end{abstract}
\thispagestyle{empty}
\vfill

\newpage

\section{Introduction}

The physics programme of the Large Hadron Collider (LHC) is centred around precision physics and will culminate in its high-luminosity phase~\cite{Azzi:2019yne}.
For the Standard Model, it amounts to comparing experimental data against precise theory predictions for a multitude of processes.
This allows to extract fundamental parameters and verify our understanding of elementary particle physics.
One key parameter is the W-boson mass as it is connected to electroweak (EW) symmetry breaking mechanism.
It is already known to a very high precision but progress in experiment and theory will allow to improve the estimate even further in the future~\cite{Azzi:2019yne}.

\begin{figure}
\centering
        \begin{subfigure}{0.49\textwidth}
                 \includegraphics[width=\textwidth]{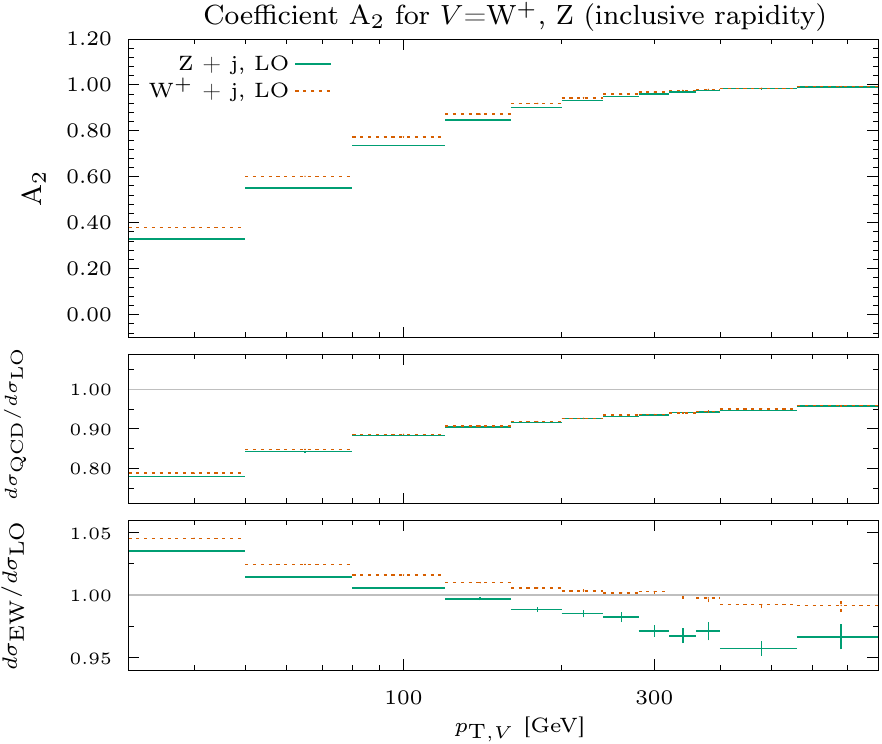}
        \end{subfigure}
        \hfill
        \begin{subfigure}{0.49\textwidth}
                 \includegraphics[width=\textwidth]{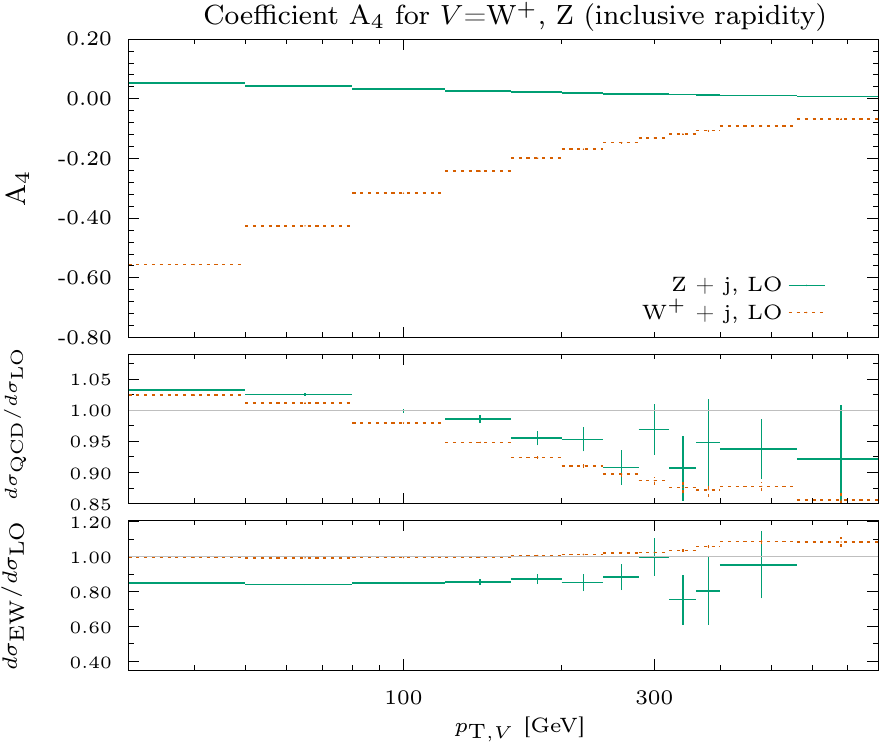}
        \end{subfigure}
        \caption{\label{fig:Z_W_comp}
                Differential distributions of the angular coefficients ${\rm A}_2$ (left) and ${\rm A}_4$ (right) for finite transverse momentum of the Z- and ${\rm W}^+$-boson. The upper panel shows the absolute LO predictions.
                The NLO QCD and NLO EW $K$-factor are displayed in the middle and lower panel, respectively. The predictions are inclusive over the whole rapidity range.
                }
\end{figure}

In that respect, at the LHC, the simplest process to investigate is the charged-current Drell-Yan process $\Pp\Pp\to\PW^\pm\to\ell^{\pm}\oset{(-)}{\nu}_{\ell}+X$.
While the undetected final state neutrino prevents direct reconstruction of the W-boson resonance, the kinematic distributions of the charged lepton carries an imprint of the W-boson mass.
The most precise mass measurements \cite{ATLAS:2017rzl,LHCb:2021bjt,CDF:2022hxs} use template fits to mass-sensitive observables like the transverse mass.
Theoretical uncertainties on the templates are a limiting factor of such measurements~\cite{ATLAS:2017rzl,LHCb:2021bjt}.
For that purpose, a particularly large source of theory uncertainty are angular decay coefficients~\cite{Strologas:2005xs,Lyu:2020nul} used to model spin correlation in W-boson decays in Monte Carlo predictions.
Currently, the coefficients are extrapolated from the $\PZ$-boson counterparts measured at the LHC~\cite{ATLAS:2017rzl}.
Higher-order perturbative corrections to the coefficients are not necessarily identical between Z- and W-boson production, as it can be seen in Fig.~\ref{fig:Z_W_comp} for the coefficients ${\rm A}_2$ and ${\rm A}_4$, defined according to Eq.~\eqref{eq:coef}.\footnote{To compute the EW corrections, the same procedure as in Ref.~\cite{Frederix:2020nyw} is followed.}

In particular, one can observe that both next-to-leading order (NLO) QCD and NLO EW corrections can differ by several per cent between the two processes.
For example, ${\rm A}_2$ show differences in the EW corrections but none for the QCD ones.
The picture is opposite for ${\rm A}_4$ where the QCD corrections are different.
In order not to rely on extrapolations from Z- to W-boson angular coefficients, we provide precise theoretical predictions for the latter with finite transverse momentum of the W-boson.

Apart from their impact on W-boson mass measurements, the angular coefficients can also be measured experimentally~\cite{Richter-Was:2016avq}.
In particular, results exist for the Tevatron at $1.8\TeV$ by the CDF collaboration~\cite{CDF:2005qwt}.
A measurement at the LHC has not been yet performed.

For its Z-boson counterpart, these coefficients are known up to next-to-next-to-leading order (NNLO) QCD~\cite{Mirkes:1994dp,Gauld:2017tww} and NLO EW~\cite{Frederix:2020nyw} accuracy.
On the experimental side, several measurements of the Z-boson angular coefficients have been performed~\cite{CDF:2011ksg,CMS:2015cyj,ATLAS:2016rnf,LHCb:2022tbc}.
In general, good agreement has been found with theoretical predictions in the Standard Model.

Notably, the production of a W-boson in association with a QCD jet is currently also known up to NNLO QCD~\cite{Boughezal:2015dva,Boughezal:2016dtm,Gehrmann-DeRidder:2019avi,Czakon:2020coa,Pellen:2021vpi}+NLO EW~\cite{Kuhn:2007qc,Kuhn:2007cv,Hollik:2007sq,Denner:2009gj} accuracy, and predictions combining QCD and EW effects have been presented in Ref.~\cite{Kallweit:2014xda,Kallweit:2015dum,Biedermann:2017yoi}.
Nonetheless, to the best of our knowledge, this accuracy is yet to be reached for W-boson angular coefficients.
We therefore aim to fill this gap by providing state-of-the-art predictions at NNLO QCD+NLO EW accuracy in the present work.

The article is organised as follows:
in section \ref{sec:details}, we define the process under investigation and what contributions are included in our computation.
In section \ref{sec:numerical_results}, the numerical results are presented and discussed.
Finally, section \ref{sec:conclusion} contains a brief summary of our findings as well as concluding remarks.

\section{Details of the calculations}\label{sec:details}

\subsection{Definition of the process}
\label{sec_definition_process}

The hadronic process of interest is 
\begin{equation}\label{eq:process}
\Pp\Pp  \rightarrow \ell^{\pm} \oset{(-)}{\nu}_{\ell}\ \Pj + \text{X} ,
\end{equation}
at the LHC.
We consider CKM to be a unity matrix as the diagonal contributions are dominant.

\paragraph{Higher-order corrections}

Using the notation scheme used in Refs.~\cite{Frederix:2018nkq} to describe the tower of contributions arising in fixed-order computations, one obtains for the process above:
\begin{align}
\label{eq:tower}
\Sigma_{\rm LO} &:= {\color{blue} \Sigma_{\text{LO}_1}}  + \Sigma_{\text{LO}_2} , \nonumber \\
\Sigma_{\rm NLO} &:=  {\color{blue} \Sigma_{\text{NLO}_1}}  + {\color{blue} \Sigma_{\text{NLO}_2}} + \Sigma_{\text{NLO}_3} , \nonumber \\
\Sigma_{\rm NNLO} &:= {\color{blue} \Sigma_{\text{NNLO}_1}} + \Sigma_{\text{NNLO}_2} + \Sigma_{\text{NNLO}_3} + \Sigma_{\text{NNLO}_4} ,
  \end{align}
where the lower index $1$ indicates the leading QCD corrections/contribution and $\Sigma$ is an infrared-safe observable.
In our case, we include the terms marked in blue that provide the leading corrections.
The other contributions are either negligible~\cite{Denner:2019zfp} or unknown.

\paragraph{QCD corrections}
The NLO and NNLO QCD corrections are computed in the 5-flavour scheme and therefore include all quark flavours apart from the top quark which is considered massive.
All partonic channels are included and the narrow-width approximation (NWA) is used.
We have verified that in the present set-up, it provides an excellent approximation of the process by comparing it against an off-shell computation at LO and NLO QCD.
In addition, in Ref.~\cite{Pellen:2021vpi} such comparisons have been performed up to NNLO QCD accuracy and the difference has been found to be negligible close to the W-boson resonance.

\begin{figure}
\centering
        \begin{subfigure}{0.32\textwidth}
                \subcaption{}
                 \includegraphics[width=\textwidth]{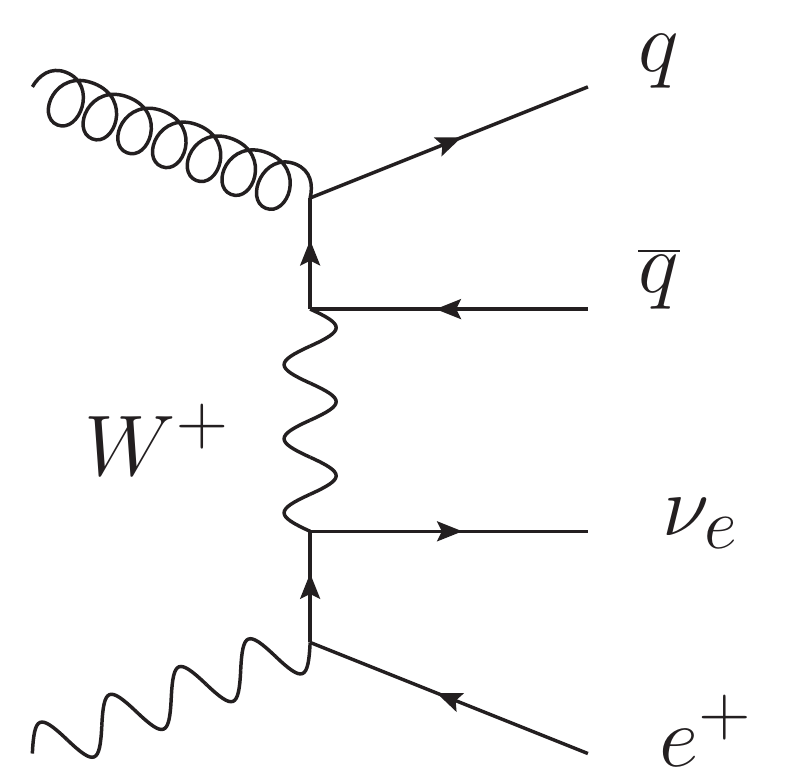}
        \end{subfigure}
        \caption{\label{fig:diag}%
                Exemplary Feynman diagram of photon-induced contributions featuring double-soft singularities at the current perturbative order. 
                }
\end{figure}

\paragraph{EW corrections}
The EW corrections contain all real photon radiation and corresponding one-loop virtual corrections, computed with all off-shell effects within the complex-mass scheme~\cite{Denner:1999gp,Denner:2005fg,Denner:2006ic}.
They also include photon induced corrections which can lead to mixing QCD-EW singularities of double soft/collinear type for some $t$-channel exchange.
An illustrative Feynman diagram of such a contribution is represented in Fig.~\ref{fig:diag}.
These divergences can be cured by applying a transverse-momentum cut on the charged leptons~\cite{Frederix:2020nyw}.
For the predictions shown in the current work, the numerical value used for the transverse momentum of the charged lepton is $1\GeV$.

\paragraph{Definition of angular coefficients}
For the definition of the angular coefficients, we follow the one used in Ref.~\cite{Frederix:2020nyw} which itself relies on Ref.~\cite{Gauld:2017tww,Mirkes:1992hu} within the Collins-Soper reference frame~\cite{Collins:1977iv}.
The expansion of the differential cross section of the process \eqref{eq:process} hence reads:
\begin{eqnarray}
\begin{split}
\label{eq:coef}
\frac{\text{d}\sigma}{\text{d}p_{{\rm T},\PW}\,\text{d}y_\PW\,\text{d}m_{\ell\nu}\,\text{d}\Omega}=
    &\frac{3}{16\pi}\frac{\text{d}\sigma^{U+L}}{\text{d}p_{{\rm T},\PW}\,\text{d}y_\PW\,\text{d}m_{\ell\nu}} \bigg((1+\cos^2\theta)+{\rm A}_0\frac{1}{2}(1-3\cos^2 \theta) \\
& +{\rm A}_1 \sin 2\theta \cos \phi + {\rm A}_2 \frac{1}{2}\sin^2\theta \cos 2\phi +{\rm A}_3\sin \theta \cos \phi +{\rm A}_4 \cos \theta \\
 & +{\rm A}_5 \sin^2 \theta \sin2\phi+{\rm A}_6 \sin2\theta \sin\phi +{\rm A}_7 \sin \theta \sin \phi\bigg) ,
\end{split}
\end{eqnarray}
where $\phi$ and $\theta$ are the azimuthal and polar angle of the charged lepton, in the Collins-Soper frame, respectively.
The cross section $\sigma^{U+L}$ represents the unpolarised cross section.
It is worth mentioning that the coefficients A$_5$--A$_7$ become non-zero starting only at order $\mathcal{O}\left(\alphas^2\right)$, justifying why these are treated separately in the following. 
Note that the definition of the angular coefficients requires the knowledge of the neutrino momentum which we assume to be accessible.

It is worth emphasising that these coefficients do not provide more information than the one contained in the invariant mass, transverse momentum, and rapidity of the lepton pair along with the rapidity and transverse momentum of one of the two leptons.
On the other hand, angular coefficients are universal pseudo-observables which allow to reduce the differential cross section to simple coefficients which factorise the production and decay process.
Moreover, the numerical determination of the coefficients is computationally easier than the 4D differential distribution in case of the leptons transverse momentum and rapidity.

In addition, while such a decomposition in Eq. \eqref{eq:coef} is exact for LO or QCD predictions, it is not for EW corrections \cite{Ebert:2020dfc}.
In this case, the photon radiation off the final-state lepton breaks this relation, as it is no longer a two-body decay.
In Appendix~\ref{sec:ref}, we show that the error induced by the radiation of final-state photons is negligible for the present practical purpose.

\paragraph{Combination}
In this work, we combine NNLO QCD and NLO EW corrections.
For a typical observable, there are usually two prescriptions that are used in the literature: the additive and the multiplicative one.
In the present case, however, this matter is slightly more complicated as we are essentially dealing with ratios.
We have opted for an unexpanded prescription as well an expansion in terms of the strong coupling which we describe in the following.

More concretely, all coefficients are written in the form
\begin{eqnarray}
\label{eq:prescr1}
{\rm A}^{\rm default}_j = \frac{N}{D} ,
\label{eq:ratio}
\end{eqnarray}
where the numerator and denominator are expanded in the strong coupling in our computations as
\begin{eqnarray}
X= \alpha_s X_1 + \alpha_s^2 X_2 + \alpha_s^3 X_3 ,
\end{eqnarray}
with $X=D,N$.
This unexpanded ratio defines our nominal prediction that we denote by \emph{def} in the rest of the article.
After expanding in terms of $\alphas$, truncating, and reordering all terms, the coefficients become
\begin{eqnarray}
\label{eq:prescr2}
{\rm A}^{\rm exp}_j = A+\alpha_s B + \alpha_s^2 C ,
\end{eqnarray}
with 
\begin{eqnarray}
\begin{split}
A &=  N_1/D_1 , \\
B&= \frac{N_2D_1-N_1D_2}{D_1^2} , \\
C&= \frac{N_3D_1^2-N_1D_3D_1+N_1D_2^2-N_2D_2D_1}{D_1^3} .
\end{split}
\end{eqnarray}
This defines our second prescription which we refer to as \emph{exp} in the rest of the article.
To determine the scale dependence we adopt an uncorrelated scale variation prescription for the renormalisation and factorisation scales in the numerator $\mu^{\text{num}}_{R/F}$ and denominator $\mu^{\text{den}}_{R/F}$, using 31-point scale variation arising from the constraint $1/2 \leq \mu^i_a / \mu^j_b \leq 2$. 

For simplicity, the EW corrections have been incorporated directly at the level of the ratio of each coefficient.
To be explicit, our prescription reads:
\begin{eqnarray}
{\rm A}_{j, {\rm QCD + EW}} = K_\text{NLO EW} \times {\rm A}_{j} ,
\end{eqnarray}
where $K_\text{NLO EW}$ is the NLO EW $K$-factor of the respective angular coefficient obtained with a cut of $1\GeV$ on the charged lepton transverse momentum, as explained above.
The coefficients ${\rm A}_{j}$ are defined in Eq.~\eqref{eq:prescr1} and \eqref{eq:prescr2} for the two prescriptions.

We would like to mention that we have explored alternative prescriptions for the ratio definition.
In particular, one can simultaneously expand Eq.~\eqref{eq:ratio} in terms of both the $\alphas$ and $\alpha$ couplings.
In this case, one obtains cross terms that one is free to include or not.
Given that one does not formally have control over these higher-order terms, we have refrained from showing results for these.
In addition, we have checked that they are anyway covered by the \emph{def} prescriptions that we present here.

\subsection{Computational set-up}

The numerical results presented here are for the LHC running at $13\TeV$.
The parton distribution functions of the protons are taken from the \texttt{LUXqed17\_plus\_nnlo} PDF set \cite{Manohar:2017eqh} which features photon PDF.
These are obtained through the LHAPDF6 program~\cite{Buckley:2014ana}.

The renormalisation and factorisation scale used is defined as the transverse energy of the lepton-neutrino pair:
\begin{eqnarray}
\mu_0 = \sqrt{m_{\ell\nu}^2+p_{\text{T},\ell\nu}^2} .
\end{eqnarray}
The EW input values are:
\begin{alignat}{2}
\label{eqn:ParticleMassesAndWidths}
                M_W &=  80.379\GeV,      & \quad \quad \quad M_Z &= 91.1876\GeV,  \nonumber \\
                \Gamma_W &=  2.085\GeV,       & \Gamma_Z &= 2.4952\GeV, 
\end{alignat}
and
\begin{eqnarray}
 G_{\mu} = 1.166380\times 10^{-5} \GeV^{-2} .
\end{eqnarray}
These parameters are used to define the electromagnetic coupling $\alpha$ in the $\bar  G_{\mu}$ scheme~\cite{Frederix:2018nkq}.

We set a cut on the lepton-neutrino transverse momentum of $p_{\text{T},\ell\nu} > 30\GeV$ in order to define the $\PW+\Pj$ process in a simple manner.
In this way the process is defined fully inclusive in the decay kinematics and no other cuts are applied to the QCD radiation in the final state.
Nonetheless, for the reason explained above, for the NLO EW corrections, a cut of $1\GeV$ is imposed on the charged lepton transverse momentum.
The leptons are dressed with final-state photons using a $R = 0.1$ cone.

In order to restrict the phase space to the resonant region, we further set a cut on the invariant mass of the lepton-neutrino pair of $m_{\ell\nu} \in [60, 100]\GeV$.
The phase-space region cut away is rather small and it also improves the agreement between the NWA and the off-shell computation.
These restrictions can simply be seen as technical cuts that do not impact the physics results.
In particular, while in the absolute predictions deviations of the order of $4\%$ can be observed (see Table~\ref{tab:inclusive}), there are no observable differences for the predictions of the coefficients given that these are normalised.

\subsection{Tools used}

The QCD corrections have been obtained from the program {\sc Stripper}, a \texttt{C++} implementation of the four-dimensional formulation of the sector-improved residue subtraction scheme \cite{Czakon:2010td,Czakon:2011ve,Czakon:2014oma,Czakon:2019tmo}.
Within this framework, $\PW+\Pc$ production~\cite{Czakon:2020coa} and polarised $\PW+\Pj$ predictions~\cite{Pellen:2021vpi} up to NNLO QCD accuracy have been previously obtained.
To that end, the {\sc AvH} library \cite{Bury:2015dla} and {\sc OpenLoops 2} \cite{Buccioni:2019sur} have been used.
In addition, two-loop amplitudes have been taken from Ref.~\cite{Gehrmann:2011ab} and numerically evaluated thanks to {\sc Ginac} \cite{Bauer:2000cp,Vollinga:2004sn}.

For the NLO electroweak corrections, we utilise the \textsc{MadGraph5\_aMC@NLO} matrix-element generator~\cite{Alwall:2014hca,Frederix:2018nkq}, in which the NLO EW corrections are fully automated.

\section{Numerical results}\label{sec:numerical_results}

In this section, we report on the theoretical predictions of the angular coefficients.
First, the inclusive cross sections are given in Table~\ref{tab:inclusive}.
The LO, NLO QCD, NNLO QCD, and NLO EW predictions in $\pb$ as well as the corresponding $K$-factors are provided for both signatures.
At LO and NLO QCD, two types of predictions are provided: a full off-shell one and one in the NWA.
As customary for $\PW+\Pj$ production, higher-order QCD corrections are rather large at NLO but moderate at NNLO QCD~\cite{Rubin:2010xp}.
In addition, the inclusion of higher-order QCD corrections is accompanied by a substantial reduction of the QCD scale uncertainty which reaches less than $3\%$ at NNLO QCD accuracy.
On the other hand, the EW corrections, driven by Sudakov logarithms~\cite{Denner:2019vbn}, are negative and grow larger in the high-energy limit.
Nonetheless, at the level of the inclusive cross section the corrections only amount to $-2\%$.
We note that these cross sections are challenging to measure experimentally due to the neutrino escaping the detectors as well as the non-tagged jet.
Nonetheless, the numbers presented should help in the reproducibility of the results presented here.

In addition, we assume throughout the article that the neutrino momentum can be fully reconstructed.
This is not the case in an experimental analysis where smearing effects have to be taken into account.
Still, the results presented here may serve as a theoretical reference, which the unfolded measurements could be compared to.

\begin{table}[h]
\begin{center}
\renewcommand{\arraystretch}{1.5}
\scriptsize
\begin{tabular}{c c c c c c c }
\toprule
Process %
     & LO $[\rm pb]$
     & NLO QCD $[\rm pb]$
     & NNLO QCD $[\rm pb]$
     & $K_{\rm NNLO}$
     & NLO EW $[\rm pb]$
     & $K_{\rm EW}$
     \\
\midrule
$\Pp\Pp \to e^- \overline{\nu}_e \Pj$
     & $896.11(6)^{+11.6\%}_{-9.4\%}$
     & $1293.4(7)^{+7.0\%}_{-6.3\%}$
     & --
     & --
     & $884.85(3)^{+17.8\%}_{-15.7\%}$
     & 0.98
     \\
\midrule
$\Pp\Pp \to e^- \overline{\nu}_e \Pj$ (NWA)
     & $928.60(3)^{+11.6\%}_{-9.4\%}$
     & $1339.5(2)^{+7.0\%}_{-6.3\%}$
     & $1448(3)^{+2.0\%}_{-2.9\%}$
     & 1.08
     & --
     & --
     \\
\midrule
$\Pp\Pp \to e^+\nu_e \Pj$
     & $1206.96(4)^{+11.6\%}_{-9.4\%}$
     & $1750(1)^{+7.1\%}_{-6.4\%}$
     & --
     & --
     & $1191.69(3)^{+17.9\%}_{-15.8\%}$
     & 0.98
     \\
\midrule
$\Pp\Pp \to e^+\nu_e \Pj$ (NWA)
     & $1250.64(4)^{+11.6\%}_{-9.4\%}$
     & $1814.9(4)^{+7.1\%}_{-6.4\%}$
     & $1960(4)^{+2.1\%}_{-2.9\%}$
     & 1.08
     & --
     & --
     \\
  \bottomrule
\end{tabular}
\end{center}
\caption{Total inclusive cross sections at LO, NLO (QCD and EW), and NNLO QCD expressed in $\pb$.
The corresponding $K$ factors are provided for NLO EW and NNLO QCD corrections.
Off-shell calculation employs an additional mass window $M_\PW \in [60, 100] \GeV$.
The EW corrections correspond to a results with a transverse momentum of $1\GeV$ on the charged lepton.
The last digits in parenthesis indicate the Monte Carlo errors.}
\label{tab:inclusive}
\end{table}

\begin{figure}
\centering
        \begin{subfigure}{0.49\textwidth}
                \subcaption{}
                 \includegraphics[width=\textwidth]{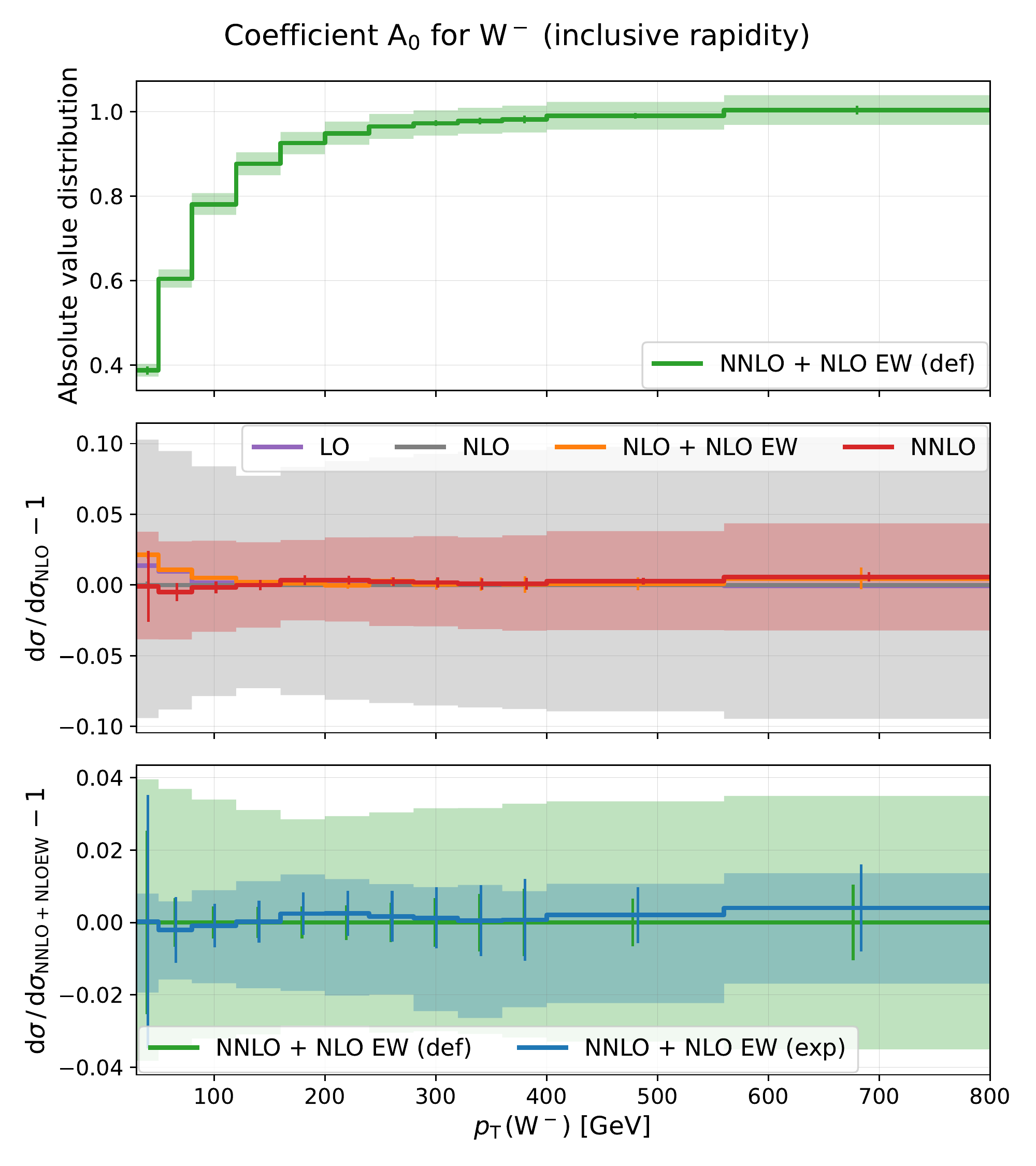}
        \end{subfigure}
        \hfill
        \begin{subfigure}{0.49\textwidth}
                \subcaption{}
                 \includegraphics[width=\textwidth]{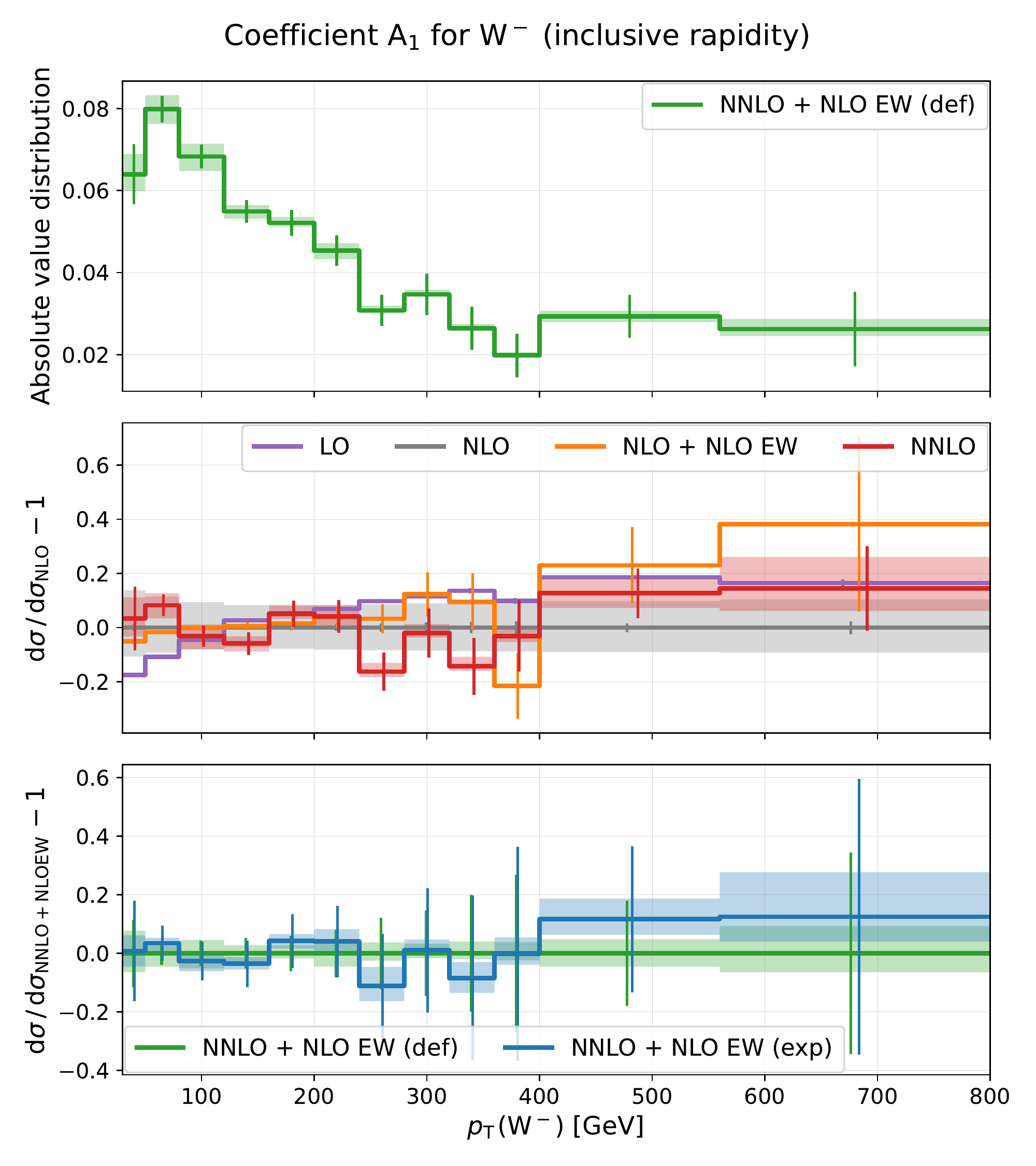}
        \end{subfigure}

        \vspace{1em}
        \begin{subfigure}{0.49\textwidth}
                \subcaption{}
                 \includegraphics[width=\textwidth]{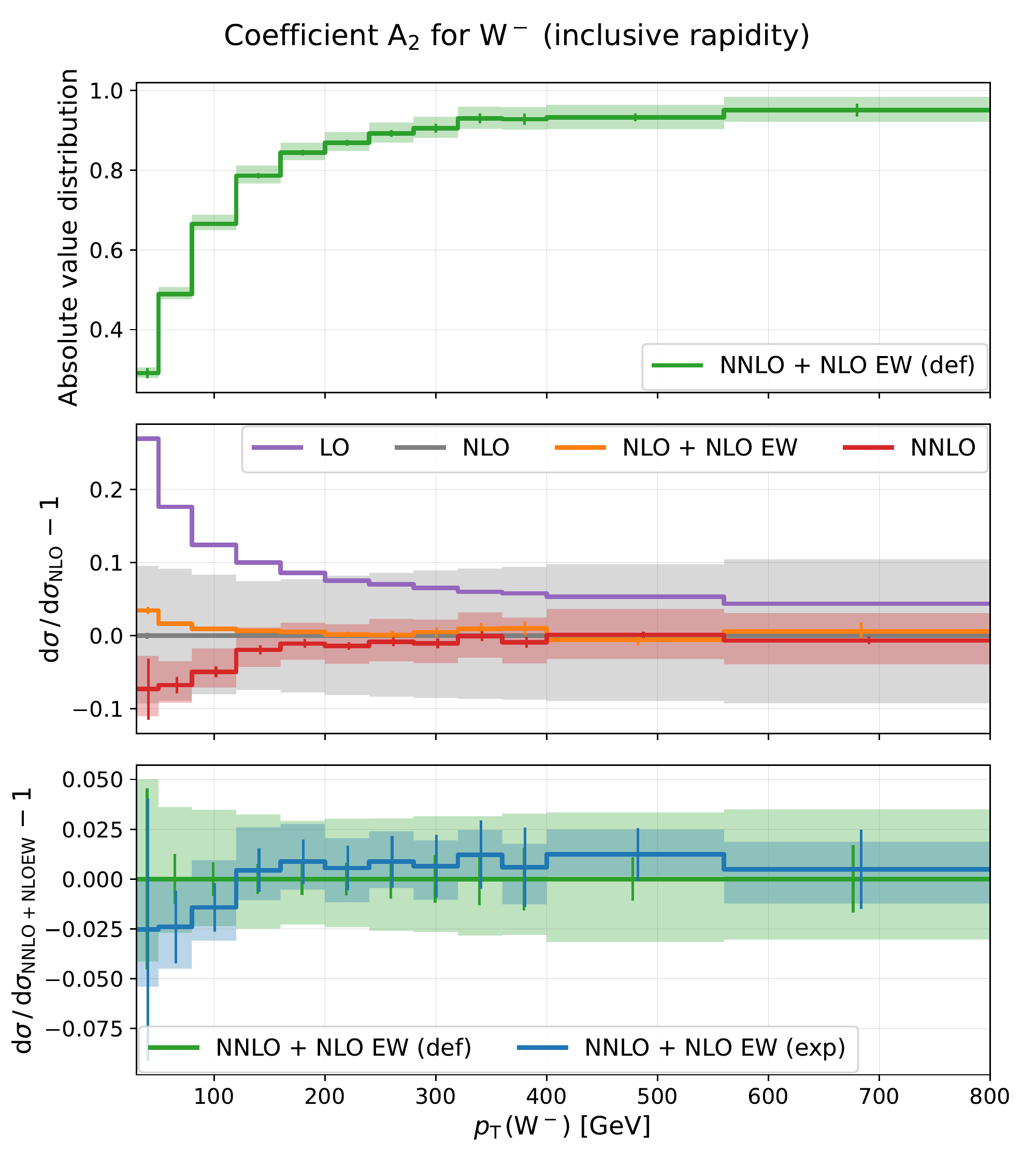}
        \end{subfigure}
        \hfill
        \begin{subfigure}{0.49\textwidth}
                \subcaption{}
                 \includegraphics[width=\textwidth]{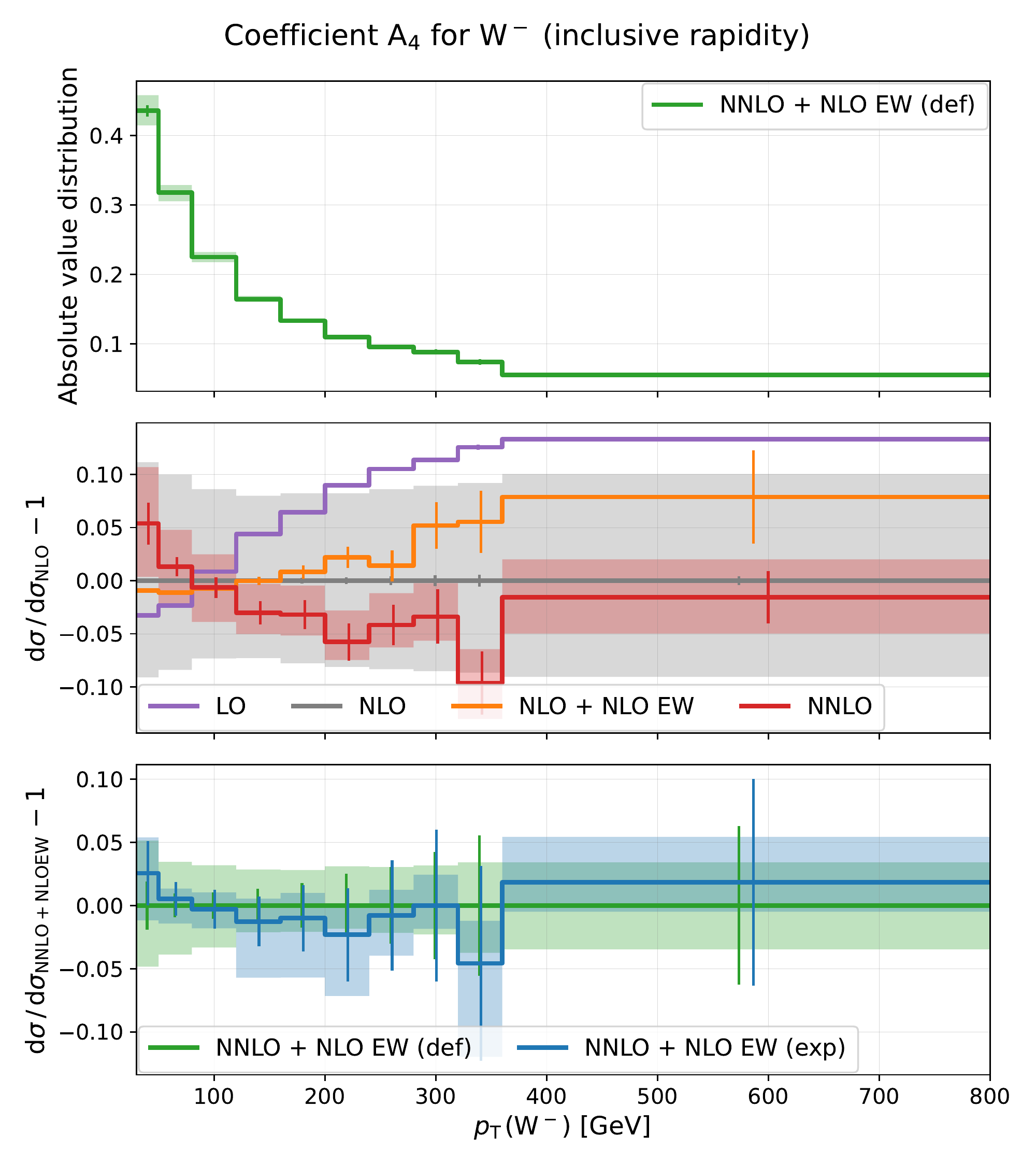}
        \end{subfigure}
        \caption{\label{fig:dist11}
                Differential distributions of the coefficient ${\rm A}_0$, ${\rm A}_1$, ${\rm A}_2$, and ${\rm A}_4$ as a function of the W-boson transverse momentum for the minus signature. 
                }
\end{figure}

\begin{figure}
\centering
        \begin{subfigure}{0.49\textwidth}
                \subcaption{}
                 \includegraphics[width=\textwidth]{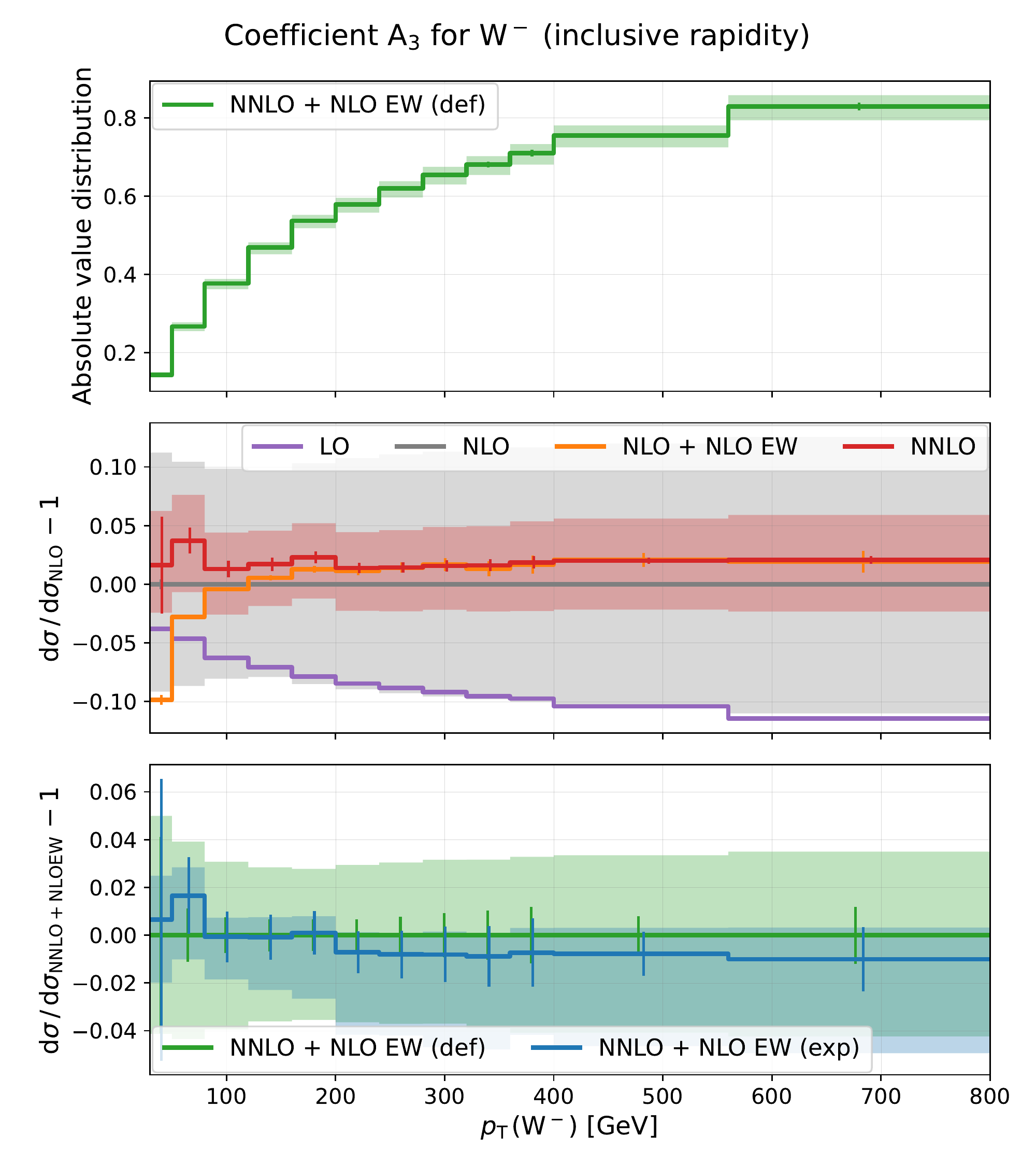}
        \end{subfigure}
        \begin{subfigure}{0.49\textwidth}
                \subcaption{}
                 \includegraphics[width=\textwidth]{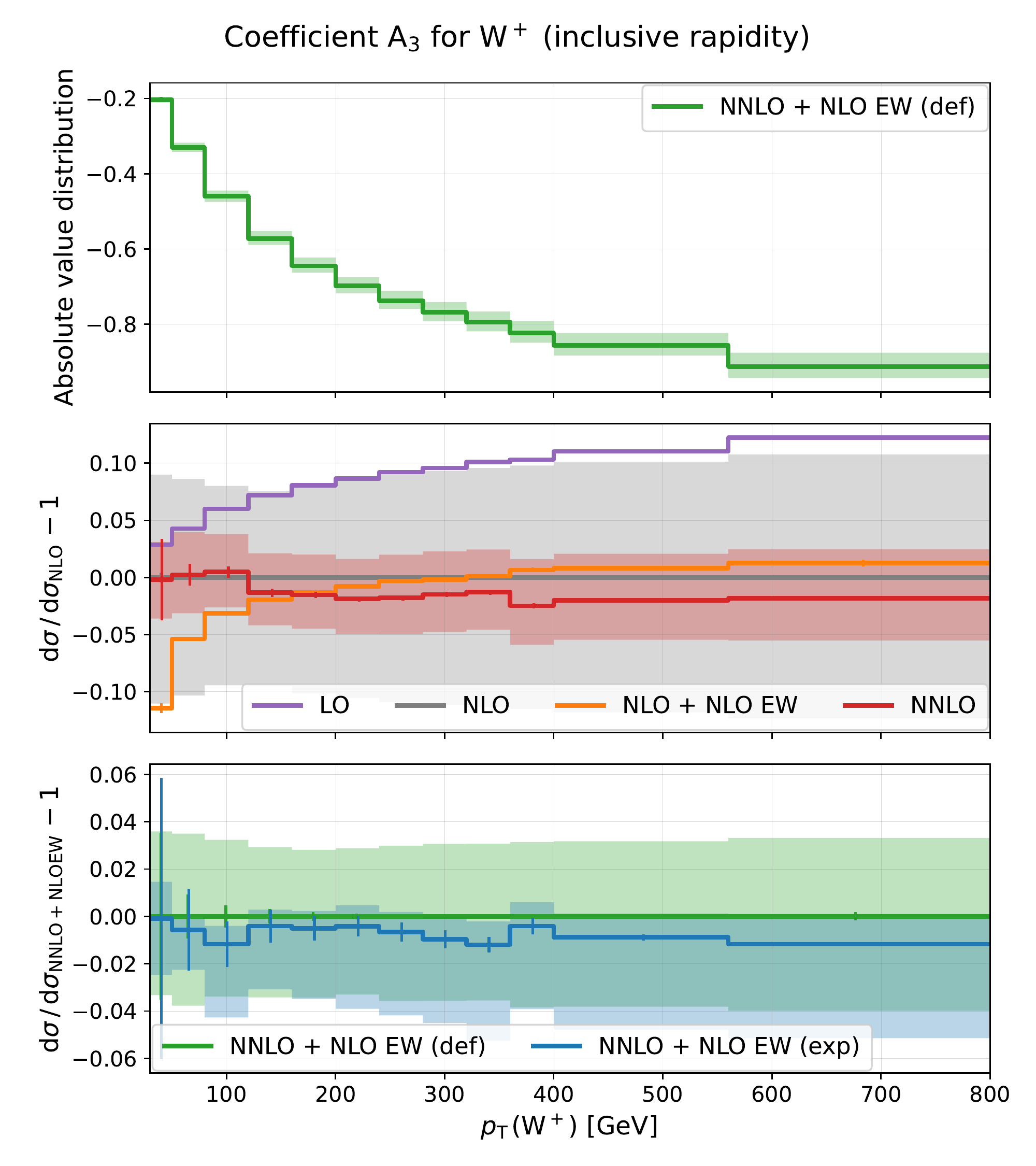}
        \end{subfigure}
        \caption{\label{fig:dist2}
                Differential distributions of the coefficient ${\rm A}_3$ as a function of the W-boson transverse momentum for the ${\rm W}^-$ (left) and ${\rm W}^+$ (right) signatures. 
                }
\end{figure}

Turning to the differential results:
we show the coefficients ${\rm A}_0$ to ${\rm A}_4$ for $\PW^-$ as a function of $p_{\rm T}$ of the W-boson in Figs.~\ref{fig:dist11} and \ref{fig:dist2} (in Fig.~\ref{fig:dist2}, the ${\rm A}_3$ coefficient is also shown for $\PW^+$).
The upper panels show our best prediction with the coefficients at NNLO QCD and NLO EW accuracy including the scale dependence.
For these predictions, the values are obtained in the default (\emph{def}) prescription defined above, see Eq.~\eqref{eq:prescr1}.
The middle panels show the different perturbative corrections with respect to NLO QCD ones.
In the lower panels the default and expanded descriptions for the ratio are compared at NNLO QCD + NLO EW accuracy.

The size of NNLO QCD and NLO EW corrections depends on the coefficient but are generally small compared to the corrections arising from NLO QCD, reproducing the observations at the level of the integrated cross section.
Apart from some exceptions which are mentioned explicitly below, the corrections are similar or identical for the $\PW^+$ and $\PW^-$ signatures.

The coefficient ${\rm A}_0$ receives flat corrections below one per cent at NNLO QCD as well as at NLO EW.
The scale dependence is reduced by a factor 2-3 when going from NLO QCD to NNLO QCD.
The expansion of the ratio does not change the central predictions but has a sizable impact on the scale dependence and reduces it by a factor of 2.

The corrections for the coefficient A${}_1$ are difficult to compute due to cancellations in the Monte Carlo integration leading to a large statistical uncertainty.
Despite the large fluctuations we can see that the corrections tend to be flat and reproducing the behaviour of A${}_0$.

For A${}_2$, NNLO QCD corrections are negative (up to $-6\%$) in particular for small transverse momentum while NLO EW corrections tend to be positive, albeit smaller.
For large transverse momentum, the corrections flatten out.
We observe a reduction of the scale dependence by a factor of 4 in this case.
The expansion prescription does induce a small difference in the shape, which, however, is still constrained within the scale dependence.

In the case of A${}_3$, NNLO QCD corrections are flat and positive at a level of roughly $2\%$.
The NLO EW correction at large transverse momentum behave similarly.
For low transverse momentum they turn negative and become larger reaching a size of $-10\%$ for $\PW^-$.
For $\PW^+$ the corrections have a similar size, but instead are positive. A direct comparison between $\PW^-$ and $\PW^+$ can be seen in Fig. \ref{fig:dist2}.
This behaviour indicates that the EW corrections are both negative (in the additive sense) for $\PW^+$ and $\PW^-$, while the NNLO QCD corrections change sign.
Some differences between the expanded and default prescription for the ratio can be observed.
While the size of the scale dependence exhibits behaviour similar to A${}_0$, there is also a shape difference reaching up to 2 per cent.

The NNLO QCD corrections for coefficient A${}_4$ are about $10\%$ and vary in shape as a function of the W-boson transverse momentum.
The NLO EW corrections are of the same order of magnitude but in opposite direction as the NNLO QCD ones.
Especially, they tend to become of the order of $7\%$ at large transverse momentum of the W boson.

\begin{figure}
\centering
        \begin{subfigure}{0.49\textwidth}
                \subcaption{}
                 \includegraphics[width=\textwidth]{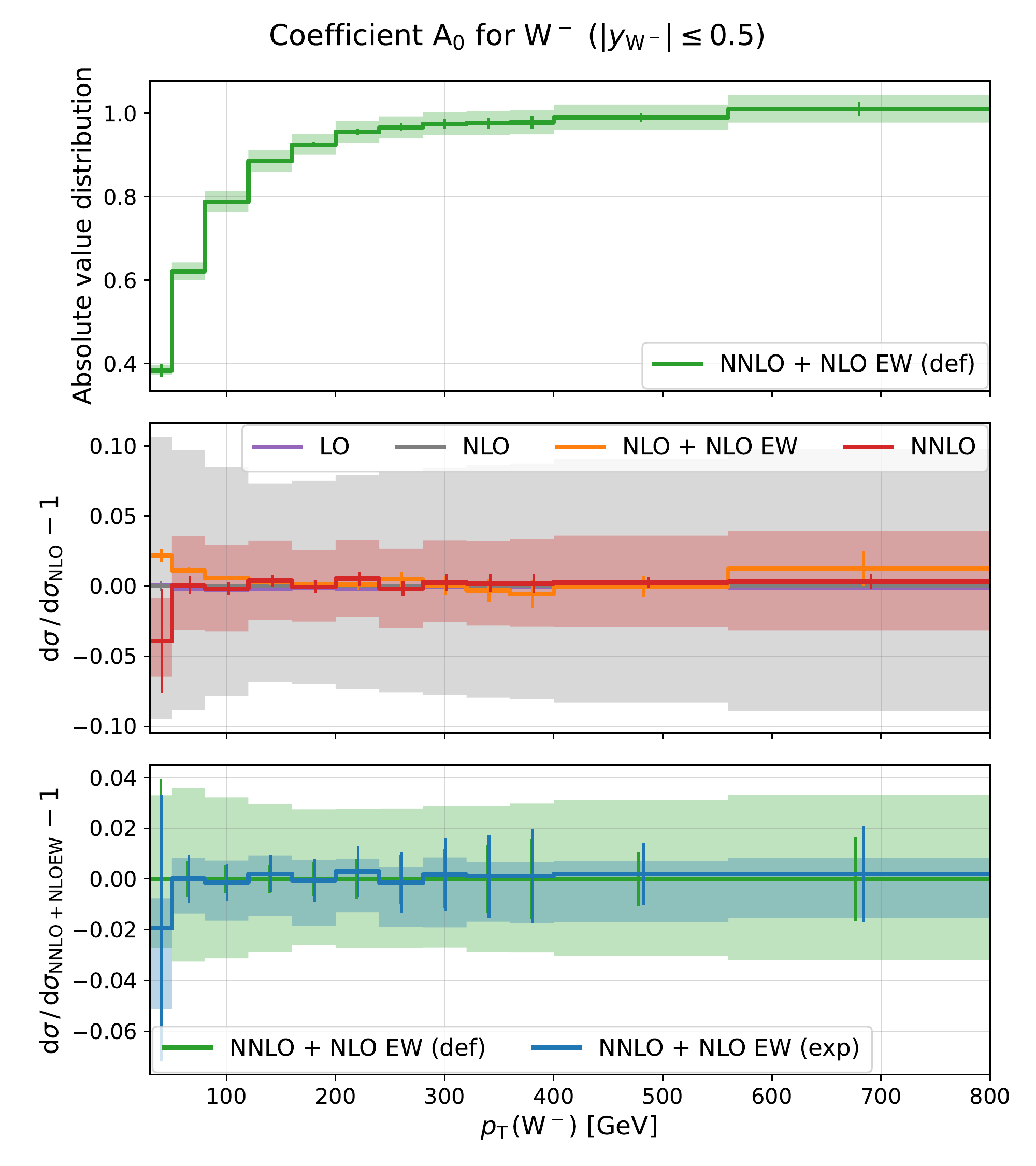}
        \end{subfigure}
        \hfill
        \begin{subfigure}{0.49\textwidth}
                \subcaption{}
                 \includegraphics[width=\textwidth]{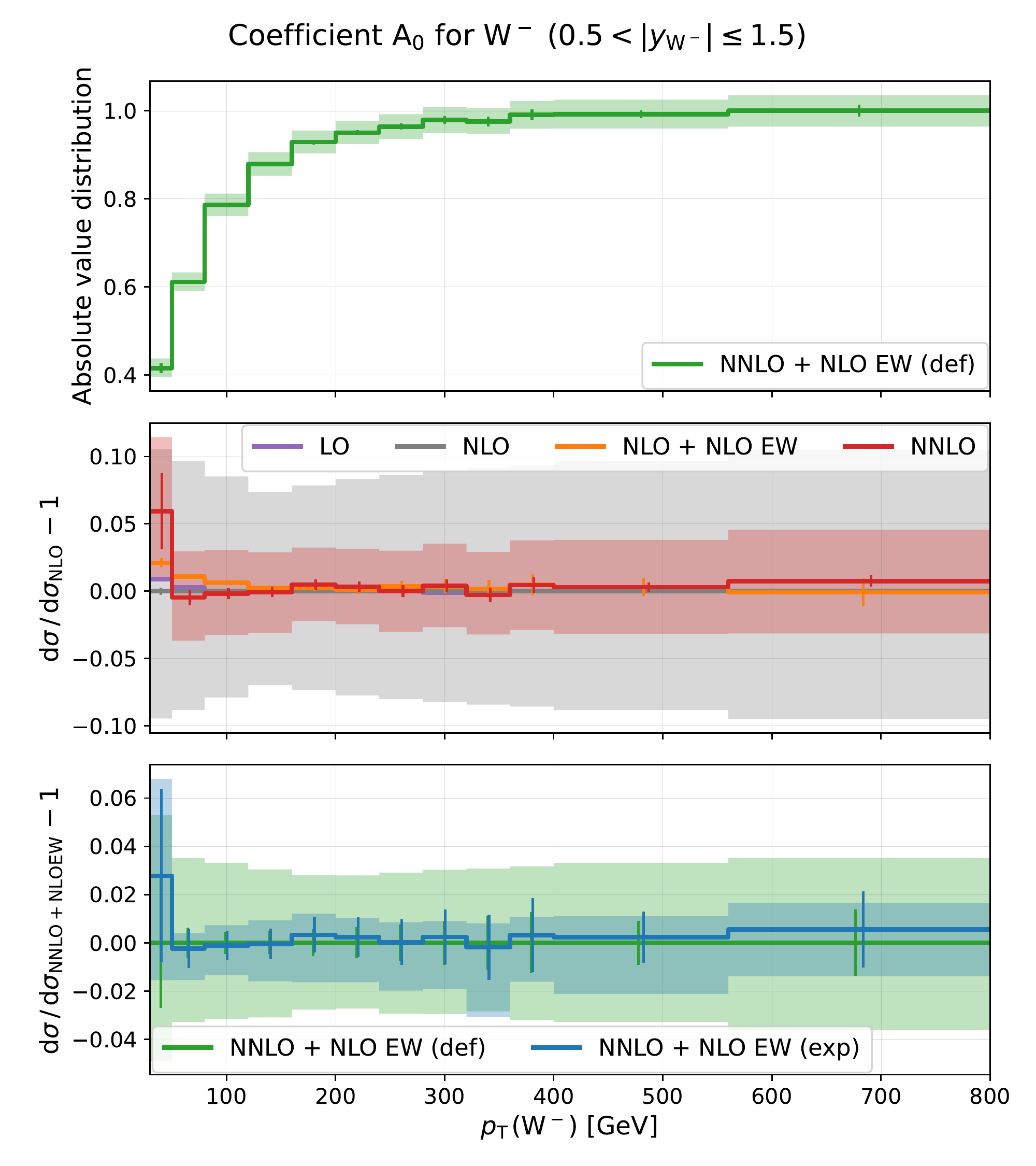}
        \end{subfigure}

        \vspace{1em}
        \begin{subfigure}{0.49\textwidth}
                \subcaption{}
                 \includegraphics[width=\textwidth]{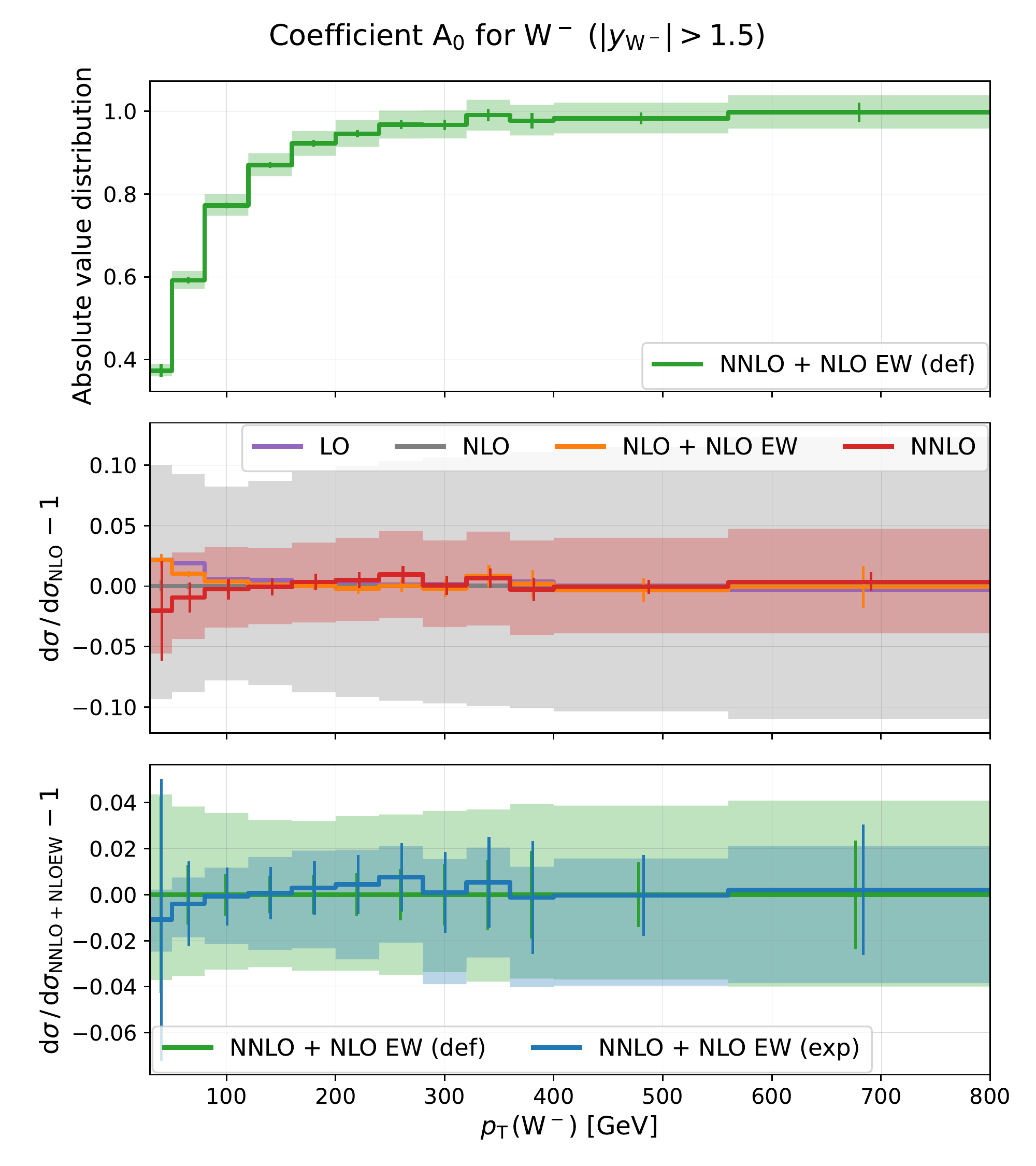}
        \end{subfigure}
        \caption{\label{fig:dist31}
                Rapidity dependence of the coefficient A$_0$ as a function of the W-boson transverse momentum. The results are shown for three rapidity bins of the W boson: $|y|\leq0.5$ (top left), $0.5<|y|\leq 1.5$ (top right), and $|y| > 1.5$ (bottom).
                }
\end{figure}

\begin{figure}
\centering
        \begin{subfigure}{0.49\textwidth}
                \subcaption{}
                 \includegraphics[width=\textwidth]{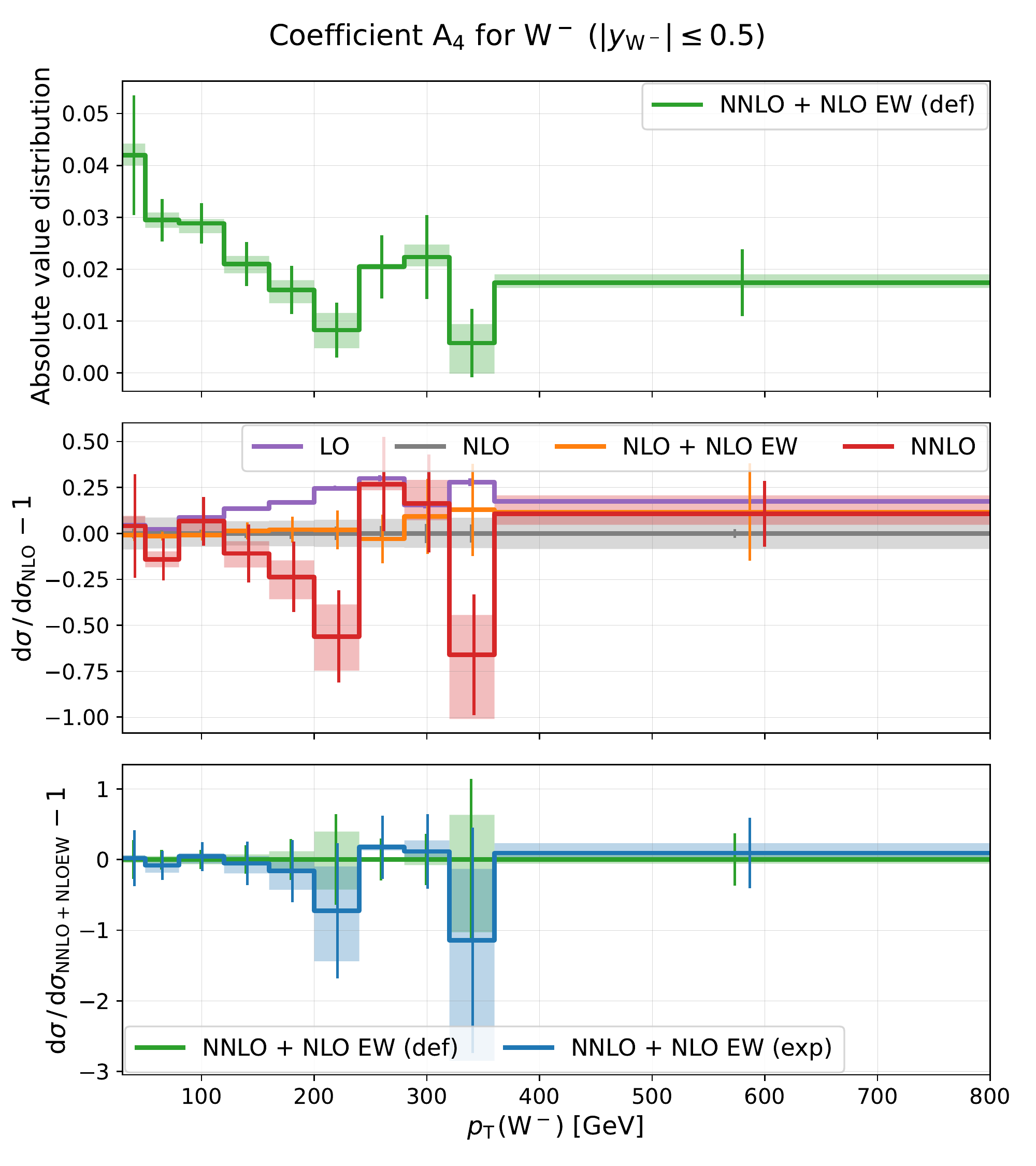}
        \end{subfigure}
        \hfill
        \begin{subfigure}{0.49\textwidth}
                \subcaption{}
                 \includegraphics[width=\textwidth]{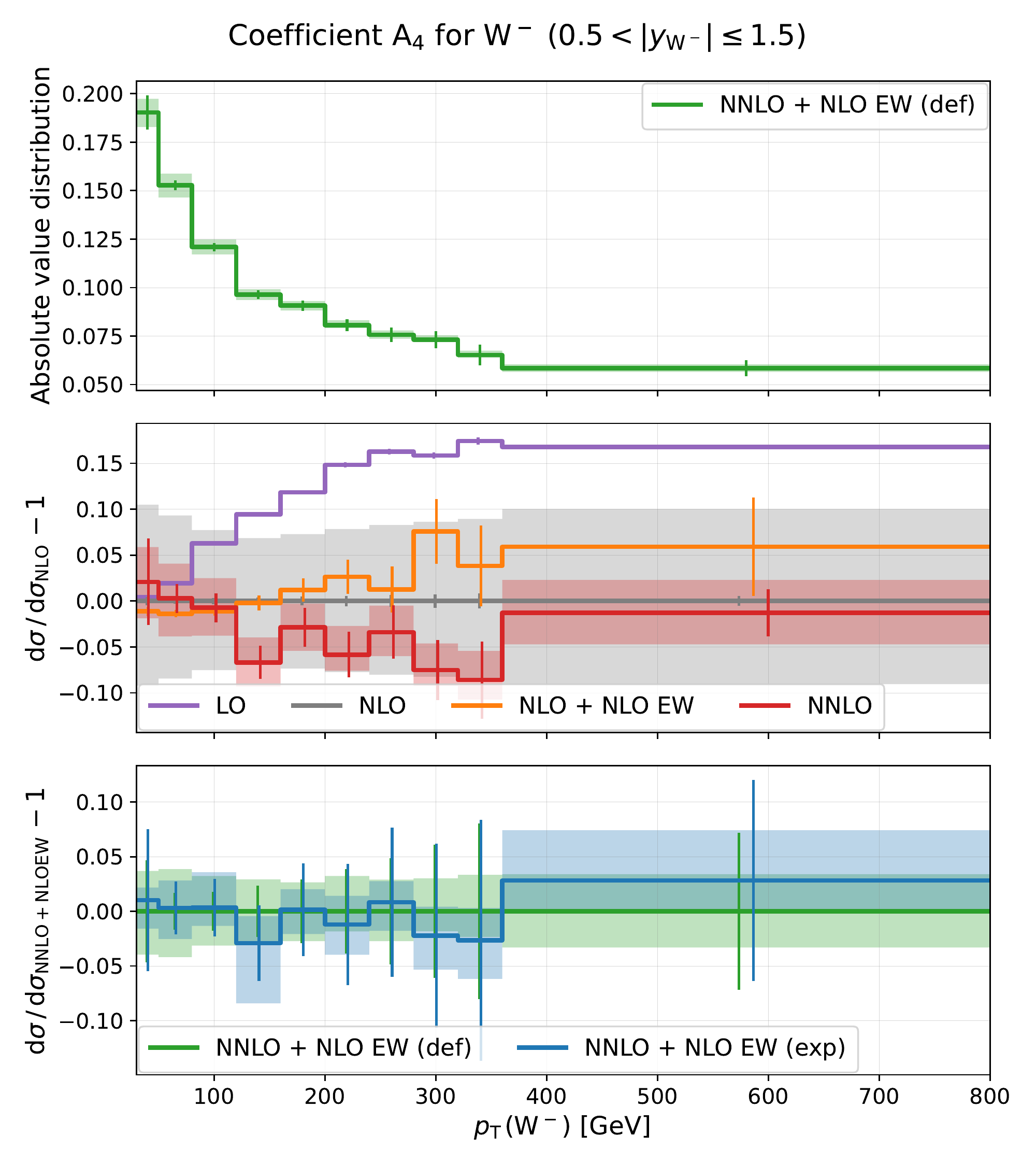}
        \end{subfigure}

        \vspace{1em}
        \begin{subfigure}{0.49\textwidth}
                \subcaption{}
                 \includegraphics[width=\textwidth]{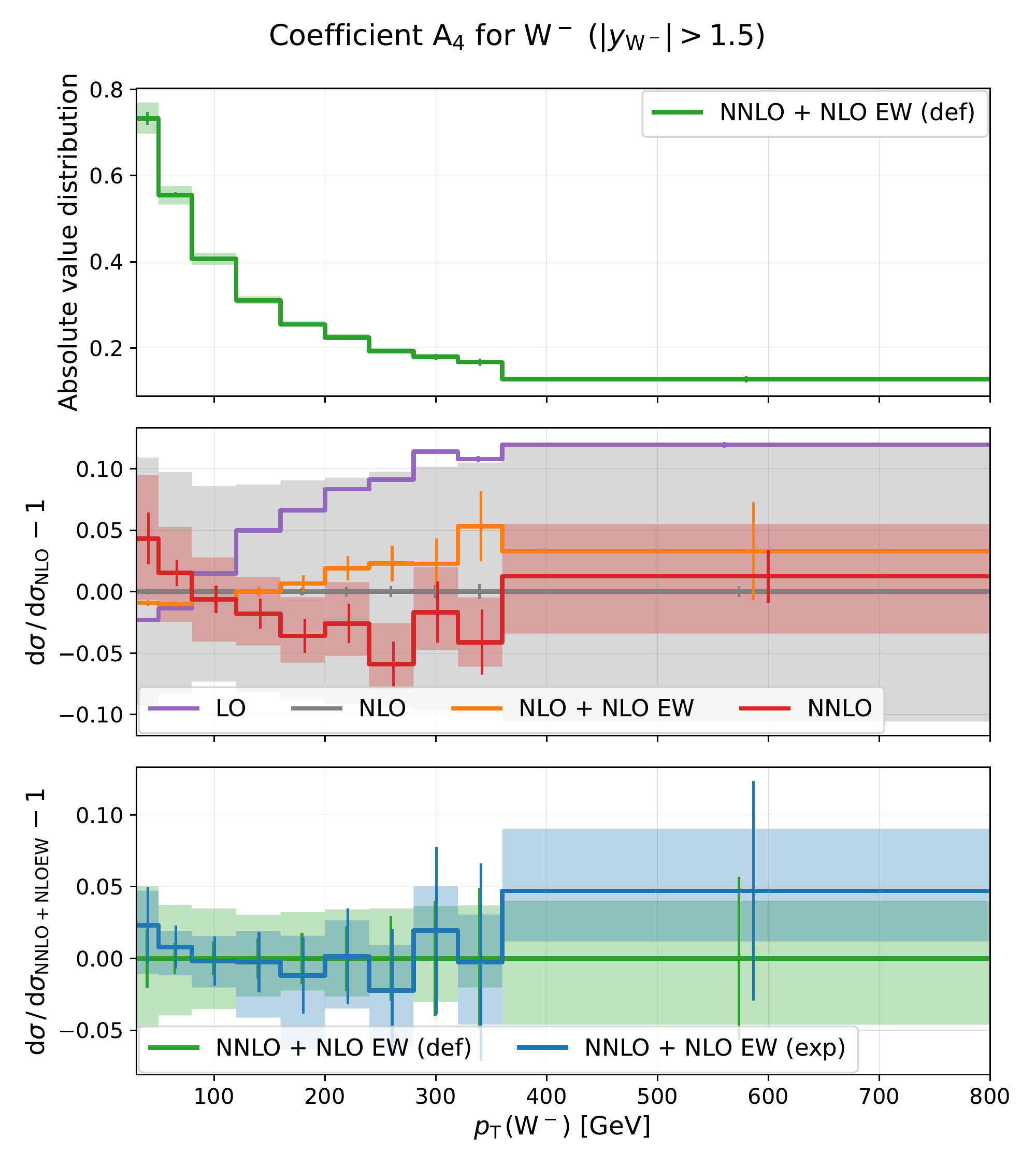}
        \end{subfigure}
        \caption{\label{fig:dist32}
                Rapidity dependence of the coefficient A$_4$ as a function of the W-boson transverse momentum. The results are shown for three rapidity bins of the W boson: $|y|\leq0.5$ (top left), $0.5<|y|\leq 1.5$ (top right), and $|y| > 1.5$ (bottom).
                }
\end{figure}

Finally, we discuss the rapidity dependence of the coefficients.
The coefficients A${}_0$ and A${}_4$ are shown in Figs.~\ref{fig:dist31} and \ref{fig:dist32} in three W-boson rapidity bins: $|y|\leq0.5$, $0.5<|y|\leq 1.5$, and $|y| > 1.5$.
The coefficient A${}_2$ behaves similar to A${}_0$, where we do not observe any significant dependence on the rapidity.
On the other hand, the coefficient A${}_4$, and similarly A${}_1$ and A$_3$ (not shown), depend strongly on the rapidity.
As shown in the lower pane of Fig.~\ref{fig:dist32}, the coefficient A${}_4$ vanishes for the central rapidity bin, which is also the reason for the large Monte Carlo fluctuations.
Hence, predictions for the bins $0.5<|y|\leq 1.5$ and $|y| > 1.5$ are significantly more stable.

We would like to mention that we have only shown results for the coefficients that are non-zero already at LO (A$_1$--A$_4$).
Nonetheless the other coefficients (A$_5$--A$_7$) are provided in the ancillary files accompanying the present submission.
These suffer from large Monte Carlo uncertainties.
Nonetheless, they might be of interest for some of the readers.


\section{Conclusion}\label{sec:conclusion}

Precision physics is one of the key ventures of the LHC, with
the study of W-boson's mass and decay properties, being an important application. In this article, we follow this avenue by computing with state-of-the-art accuracy the decay coefficients of the W-boson as a function of the W-boson transverse momentum.

For the first time, the W-boson decay coefficients are presented with NNLO~QCD and NLO~EW accuracy in $\PW+\Pj$ production at the LHC, \emph{i.e.}\ for finite transverse momentum of the W-boson, both for the W$^+$ and W$^-$ signatures.
We have found that the NNLO QCD corrections can reach $5\%$ and reduce the scale dependence by a factor of 2--4 with respect to the NLO~QCD predictions.
The EW corrections display effects of the same size and are therefore equally important.
The findings presented in this article are robust under re-expanding the coefficients in $\alpha_{\rm s}$, showing therefore reliable behaviours.

Finally, we would like to emphasize that the predictions obtained in this work are available as ancillary files in the present submission. 
In particular, the results for all coefficients are provided for the two prescriptions of our best predictions at NNLO~QCD+NLO~EW accuracy.
In this way, this work provides all necessary theoretical ingredients for accurate extraction of the angular coefficients in experimental analyses and hence contributes to the precision programme of the LHC.

\section*{Acknowledgements}

The authors would like to thank Rikkert Frederix for his contribution and reviewing the manuscript.
The authors would like to particularly thank Micha\l{} Czakon for making the {\sc Stripper} library available to us.
This research has received funding from the European Research
Council (ERC) under the European Union’s Horizon 2020 Research and Innovation
Programme (grant agreement no.\ 683211).
M.P.\ acknowledges support by the German Research Foundation (DFG) through the Research Training Group RTG2044.
A.P.\ is also supported by the Cambridge Trust and Trinity College Cambridge.
R.P.\ acknowledges the support from the Leverhulme Trust and the Isaac Newton Trust,
as well as the use of the DiRAC Cumulus HPC facility under Grant No. PPSP226.
T.V.\ is supported by the Swedish Research Council under contract number 2016-05996.

\appendix

\section{Electroweak corrections and decay coefficients}
\label{sec:ref}

As mentioned in Section~\ref{sec_definition_process}, Eq.~\eqref{eq:coef} does not strictly hold when considering EW corrections.
In particular, the radiation of photons off the final-state lepton induces a three-body decay which is not covered by the formula.
This analysis therefore raises the question as to what extent the use of Eq.~\eqref{eq:coef} is justified when computing EW corrections.

\begin{figure}
\centering
        \begin{subfigure}{\textwidth}
                \subcaption{}
                 \includegraphics[width=\textwidth]{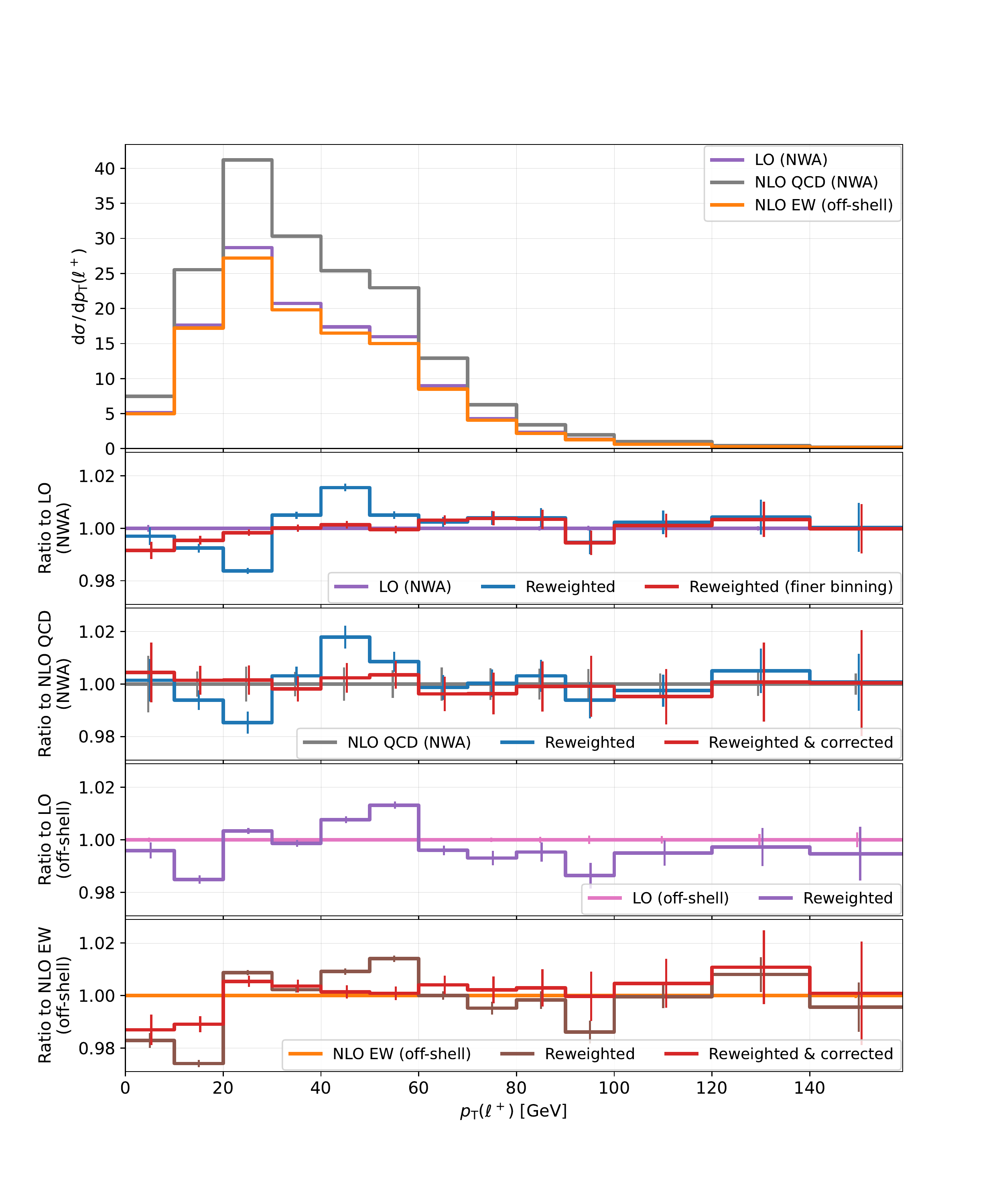}
        \end{subfigure}
        \vspace{-1.2cm}
        \caption{\label{fig:appendix}
                Transverse momentum distribution of the positron.
                The upper plot shows the absolute predictions at LO, NLO QCD, and NLO EW accuracy.
                The first inset is the comparison of the LO prediction in the NWA approximation against on-shell W-boson reweighted with the angular coefficients.
                The second inset is at NLO QCD accuracy, considering for the reweighting an additional correction taken from the LO result.
                The third inset is the same as the first one but instead of using the NWA approximation, an off-shell computation is used.
                The lowest inset is similar to NLO QCD but for NLO EW corrections.
                }
\end{figure}

One way of testing this is by checking that the full lepton distributions can be reproduced using the decay coefficients.
To this end, we generated unweighted W+j events with on-shell W-bosons, which we further reweighted using the angular coefficients in the manner prescribed by Eq.~\eqref{eq:coef}.
If the reweighted results end up close to the original distribution, we can conclude that the assumption, that EW corrections can also be described by Eq.~\eqref{eq:coef}, is valid for practical purposes.
Our results for the transverse momentum distribution are shown in Fig.~\ref{fig:appendix},
where the upper part features absolute values for differential distributions at LO, NLO QCD, and NLO EW accuracy,
and the lower part is comprised of four insets showing various ratios discussed below.

The first inset is the comparison of the LO prediction in the NWA approximation against the on-shell W-boson calculation reweighted with the angular coefficients.
The reweighting procedure works quite well, but shows some few per cent differences.
Their origin is the size of the binning used for angular coefficients, as is shown by another distribution reweighted with angular coefficients calculated with finer binning, where the agreement with the original distribution is improved.

A similar comparison is then carried out at NLO QCD accuracy, which results are presented in the second inset.
In addition to reweighting of the events with the angular coefficients, we account for the overall normalisation of the NLO QCD cross section, \emph{i.e.} the K-factor for the production part.
The fluctuations feature a very similar shape to the distribution in the inset above,
especially in the low transverse-momentum region.
Therefore, we applied a correction for the binning effects using a comparison at LO above.
The corrected results show a significant improvement reaching the level of agreement similar to the LO calculation with a finer binning.

The third inset considers the LO case, similar to the first inset, but the comparison is made against an off-shell computation instead.
In addition, the reweighting procedure further includes the off-shell/NWA factor to account for the missing off-shell effects.
Similarly to the NWA case, we observe small fluctuations present at the per cent level.
We attribute the disagreement to the binning effect, and extract the fraction into a correction factor to use further at NLO EW.

Finally, the last inset features the NLO EW corrections.
As for the other cases, the fluctuations are found at the per cent level.
Following the comparison at LO, the results are corrected for the binning effects.
The corrected results then show an improved agreement with the NLO EW computation.
We note that the first two bins show roughly two-sigma discrepancies at the level of about one per cent.
This should be taken as the residual uncertainty of the method to use the angular coefficients decay decomposition for EW corrections.
We also note that this uncertainty does not directly translate into the calculation of the NLO EW angular coefficients.
Overall, this demonstrates that the use of decay coefficients with NLO EW corrections is justified for practical purpose.

We remark that the presented results consider only the lepton transverse momentum.
A similar exercise could be performed for its rapidity distribution.
In this case, we recommend using a fine binning and extending the angular coefficients distribution range to cover higher rapidities.
Otherwise, a non-trivial rapidity dependence featured by some coefficients is not captured well and leands to a systematic bias in the result.

\bibliographystyle{utphys.bst}
\bibliography{w_coeff}

\providecommand{\href}[2]{#2}\begingroup\raggedright\begin{thebibliography}{10}

\bibitem{Azzi:2019yne}
P.~Azzi {\em et al.}, {\em {Report from Working Group 1}: {Standard Model
  Physics at the HL-LHC and HE-LHC}}.
  \href{http://dx.doi.org/10.23731/CYRM-2019-007.1}{CERN Yellow Rep. Monogr.
  {\bf 7} (2019)  1--220}, \href{http://arxiv.org/abs/1902.04070}{{\tt
  arXiv:1902.04070 [hep-ph]}}.

\bibitem{ATLAS:2017rzl}
{\bf ATLAS} Collaboration, M.~Aaboud {\em et al.}, {\em {Measurement of the
  $W$-boson mass in pp collisions at $\sqrt{s}=7$ TeV with the ATLAS
  detector}}. \href{http://dx.doi.org/10.1140/epjc/s10052-017-5475-4}{Eur.
  Phys. J. C {\bf 78} (2018) no.~2, 110},
  \href{http://arxiv.org/abs/1701.07240}{{\tt arXiv:1701.07240 [hep-ex]}}.
  [Erratum: Eur.Phys.J.C 78, 898 (2018)].

\bibitem{LHCb:2021bjt}
{\bf LHCb} Collaboration, R.~Aaij {\em et al.}, {\em {Measurement of the W
  boson mass}}. \href{http://dx.doi.org/10.1007/JHEP01(2022)036}{JHEP {\bf 01}
  (2022)  036}, \href{http://arxiv.org/abs/2109.01113}{{\tt arXiv:2109.01113
  [hep-ex]}}.

\bibitem{CDF:2022hxs}
{\bf CDF} Collaboration, T.~Aaltonen {\em et al.}, {\em {High-precision
  measurement of the W boson mass with the CDF II detector}}.
  \href{http://dx.doi.org/10.1126/science.abk1781}{Science {\bf 376} (2022)
  no.~6589, 170--176}.

\bibitem{Strologas:2005xs}
J.~Strologas and S.~Errede, {\em {Study of the angular coefficients and
  corresponding helicity cross sections of the W boson in hadron collisions}}.
  \href{http://dx.doi.org/10.1103/PhysRevD.73.052001}{Phys. Rev. D {\bf 73}
  (2006)  052001}, \href{http://arxiv.org/abs/hep-ph/0503291}{{\tt
  arXiv:hep-ph/0503291}}.

\bibitem{Lyu:2020nul}
Y.~Lyu, W.-C. Chang, R.~E. Mcclellan, J.-C. Peng, and O.~Teryaev, {\em {Lepton
  angular distribution of $W$ boson productions}}.
  \href{http://dx.doi.org/10.1103/PhysRevD.103.034011}{Phys. Rev. D {\bf 103}
  (2021) no.~3, 034011}, \href{http://arxiv.org/abs/2010.01826}{{\tt
  arXiv:2010.01826 [hep-ph]}}.

\bibitem{Frederix:2020nyw}
R.~Frederix and T.~Vitos, {\em {Electroweak corrections to the angular
  coefficients in finite-$p_T$Z-boson production and dilepton decay}}.
  \href{http://dx.doi.org/10.1140/epjc/s10052-020-08513-7}{Eur. Phys. J. C {\bf
  80} (2020) no.~10, 939}, \href{http://arxiv.org/abs/2007.08867}{{\tt
  arXiv:2007.08867 [hep-ph]}}.

\bibitem{Richter-Was:2016avq}
E.~Richter-Was and Z.~Was, {\em {W production at LHC: lepton angular
  distributions and reference frames for probing hard QCD}}.
  \href{http://dx.doi.org/10.1140/epjc/s10052-017-4649-4}{Eur. Phys. J. C {\bf
  77} (2017) no.~2, 111}, \href{http://arxiv.org/abs/1609.02536}{{\tt
  arXiv:1609.02536 [hep-ph]}}.

\bibitem{CDF:2005qwt}
{\bf CDF} Collaboration, D.~Acosta {\em et al.}, {\em {Measurement of the
  azimuthal angle distribution of leptons from $W$ boson decays as a function
  of the $W$ transverse momentum in $p \bar{p}$ collisions at $\sqrt{s} = 1.8$
  TeV}}. \href{http://dx.doi.org/10.1103/PhysRevD.73.052002}{Phys. Rev. D {\bf
  73} (2006)  052002}, \href{http://arxiv.org/abs/hep-ex/0504020}{{\tt
  arXiv:hep-ex/0504020}}.

\bibitem{Mirkes:1994dp}
E.~Mirkes and J.~Ohnemus, {\em {Angular distributions of Drell-Yan lepton pairs
  at the Tevatron: Order $\alpha-s^{2}$ corrections and Monte Carlo studies}}.
  \href{http://dx.doi.org/10.1103/PhysRevD.51.4891}{Phys. Rev. D {\bf 51}
  (1995)  4891--4904}, \href{http://arxiv.org/abs/hep-ph/9412289}{{\tt
  arXiv:hep-ph/9412289}}.

\bibitem{Gauld:2017tww}
R.~Gauld, A.~Gehrmann-De~Ridder, T.~Gehrmann, E.~W.~N. Glover, and A.~Huss,
  {\em {Precise predictions for the angular coefficients in Z-boson production
  at the LHC}}. \href{http://dx.doi.org/10.1007/JHEP11(2017)003}{JHEP {\bf 11}
  (2017)  003}, \href{http://arxiv.org/abs/1708.00008}{{\tt arXiv:1708.00008
  [hep-ph]}}.

\bibitem{CDF:2011ksg}
{\bf CDF} Collaboration, T.~Aaltonen {\em et al.}, {\em {First Measurement of
  the Angular Coefficients of Drell-Yan $e^{+}e^{-}$ pairs in the Z Mass Region
  from $p\bar{p}$ Collisions at $\sqrt{s}$ = 1.96 TeV}}.
  \href{http://dx.doi.org/10.1103/PhysRevLett.106.241801}{Phys. Rev. Lett. {\bf
  106} (2011)  241801}, \href{http://arxiv.org/abs/1103.5699}{{\tt
  arXiv:1103.5699 [hep-ex]}}.

\bibitem{CMS:2015cyj}
{\bf CMS} Collaboration, V.~Khachatryan {\em et al.}, {\em {Angular
  coefficients of Z bosons produced in pp collisions at $\sqrt{s}$ = 8 TeV and
  decaying to $\mu^+ \mu^-$ as a function of transverse momentum and
  rapidity}}. \href{http://dx.doi.org/10.1016/j.physletb.2015.08.061}{Phys.
  Lett. B {\bf 750} (2015)  154--175},
  \href{http://arxiv.org/abs/1504.03512}{{\tt arXiv:1504.03512 [hep-ex]}}.

\bibitem{ATLAS:2016rnf}
{\bf ATLAS} Collaboration, G.~Aad {\em et al.}, {\em {Measurement of the
  angular coefficients in $Z$-boson events using electron and muon pairs from
  data taken at $\sqrt{s}=8$ TeV with the ATLAS detector}}.
  \href{http://dx.doi.org/10.1007/JHEP08(2016)159}{JHEP {\bf 08} (2016)  159},
  \href{http://arxiv.org/abs/1606.00689}{{\tt arXiv:1606.00689 [hep-ex]}}.

\bibitem{LHCb:2022tbc}
{\bf LHCb} Collaboration, R.~Aaij {\em et al.}, {\em {First measurement of the
  $Z\rightarrow \mu^+ \mu^-$ angular coefficients in the forward region of $pp$
  collisions at $\sqrt{s}=13$ TeV}}.
  \href{http://arxiv.org/abs/2203.01602}{{\tt arXiv:2203.01602 [hep-ex]}}.

\bibitem{Boughezal:2015dva}
R.~Boughezal, C.~Focke, X.~Liu, and F.~Petriello, {\em {$W$-boson production in
  association with a jet at next-to-next-to-leading order in perturbative
  QCD}}. \href{http://dx.doi.org/10.1103/PhysRevLett.115.062002}{Phys. Rev.
  Lett. {\bf 115} (2015) no.~6, 062002},
  \href{http://arxiv.org/abs/1504.02131}{{\tt arXiv:1504.02131 [hep-ph]}}.

\bibitem{Boughezal:2016dtm}
R.~Boughezal, X.~Liu, and F.~Petriello, {\em {W-boson plus jet differential
  distributions at NNLO in QCD}}.
  \href{http://dx.doi.org/10.1103/PhysRevD.94.113009}{Phys. Rev. D {\bf 94}
  (2016) no.~11, 113009}, \href{http://arxiv.org/abs/1602.06965}{{\tt
  arXiv:1602.06965 [hep-ph]}}.

\bibitem{Gehrmann-DeRidder:2019avi}
A.~Gehrmann-De~Ridder, T.~Gehrmann, E.~W.~N. Glover, A.~Huss, and D.~M. Walker,
  {\em {Vector Boson Production in Association with a Jet at Forward
  Rapidities}}. \href{http://dx.doi.org/10.1140/epjc/s10052-019-7010-2}{Eur.
  Phys. J. C {\bf 79} (2019) no.~6, 526},
  \href{http://arxiv.org/abs/1901.11041}{{\tt arXiv:1901.11041 [hep-ph]}}.

\bibitem{Czakon:2020coa}
M.~Czakon, A.~Mitov, M.~Pellen, and R.~Poncelet, {\em {NNLO QCD predictions for
  W+c-jet production at the LHC}}.
  \href{http://dx.doi.org/10.1007/JHEP06(2021)100}{JHEP {\bf 06} (2021)  100},
  \href{http://arxiv.org/abs/2011.01011}{{\tt arXiv:2011.01011 [hep-ph]}}.

\bibitem{Pellen:2021vpi}
M.~Pellen, R.~Poncelet, and A.~Popescu, {\em {Polarised W+j production at the
  LHC: a study at NNLO QCD accuracy}}.
  \href{http://dx.doi.org/10.1007/JHEP02(2022)160}{JHEP {\bf 02} (2022)  160},
  \href{http://arxiv.org/abs/2109.14336}{{\tt arXiv:2109.14336 [hep-ph]}}.

\bibitem{Kuhn:2007qc}
J.~H. Kuhn, A.~Kulesza, S.~Pozzorini, and M.~Schulze, {\em {Electroweak
  corrections to large transverse momentum production of W bosons at the LHC}}.
  \href{http://dx.doi.org/10.1016/j.physletb.2007.06.028}{Phys. Lett. B {\bf
  651} (2007)  160--165}, \href{http://arxiv.org/abs/hep-ph/0703283}{{\tt
  arXiv:hep-ph/0703283}}.

\bibitem{Kuhn:2007cv}
J.~H. Kuhn, A.~Kulesza, S.~Pozzorini, and M.~Schulze, {\em {Electroweak
  corrections to hadronic production of W bosons at large transverse momenta}}.
  \href{http://dx.doi.org/10.1016/j.nuclphysb.2007.12.029}{Nucl. Phys. B {\bf
  797} (2008)  27--77}, \href{http://arxiv.org/abs/0708.0476}{{\tt
  arXiv:0708.0476 [hep-ph]}}.

\bibitem{Hollik:2007sq}
W.~Hollik, T.~Kasprzik, and B.~A. Kniehl, {\em {Electroweak corrections to
  W-boson hadroproduction at finite transverse momentum}}.
  \href{http://dx.doi.org/10.1016/j.nuclphysb.2007.09.013}{Nucl. Phys. B {\bf
  790} (2008)  138--159}, \href{http://arxiv.org/abs/0707.2553}{{\tt
  arXiv:0707.2553 [hep-ph]}}.

\bibitem{Denner:2009gj}
A.~Denner, S.~Dittmaier, T.~Kasprzik, and A.~Muck, {\em {Electroweak
  corrections to W + jet hadroproduction including leptonic W-boson decays}}.
  \href{http://dx.doi.org/10.1088/1126-6708/2009/08/075}{JHEP {\bf 08} (2009)
  075}, \href{http://arxiv.org/abs/0906.1656}{{\tt arXiv:0906.1656 [hep-ph]}}.

\bibitem{Kallweit:2014xda}
S.~Kallweit, J.~M. Lindert, P.~Maierh\"ofer, S.~Pozzorini, and M.~Sch\"onherr,
  {\em {NLO electroweak automation and precise predictions for W+multijet
  production at the LHC}}.
  \href{http://dx.doi.org/10.1007/JHEP04(2015)012}{JHEP {\bf 04} (2015)  012},
  \href{http://arxiv.org/abs/1412.5157}{{\tt arXiv:1412.5157 [hep-ph]}}.

\bibitem{Kallweit:2015dum}
S.~Kallweit, J.~M. Lindert, P.~Maierhofer, S.~Pozzorini, and M.~Sch\"onherr,
  {\em {NLO QCD+EW predictions for V + jets including off-shell vector-boson
  decays and multijet merging}}.
  \href{http://dx.doi.org/10.1007/JHEP04(2016)021}{JHEP {\bf 04} (2016)  021},
  \href{http://arxiv.org/abs/1511.08692}{{\tt arXiv:1511.08692 [hep-ph]}}.

\bibitem{Biedermann:2017yoi}
B.~Biedermann, S.~Br\"auer, A.~Denner, M.~Pellen, S.~Schumann, and J.~M.
  Thompson, {\em {Automation of NLO QCD and EW corrections with Sherpa and
  Recola}}. \href{http://dx.doi.org/10.1140/epjc/s10052-017-5054-8}{Eur. Phys.
  J. C {\bf 77} (2017)  492}, \href{http://arxiv.org/abs/1704.05783}{{\tt
  arXiv:1704.05783 [hep-ph]}}.

\bibitem{Frederix:2018nkq}
R.~Frederix, S.~Frixione, V.~Hirschi, D.~Pagani, H.~S. Shao, and M.~Zaro, {\em
  {The automation of next-to-leading order electroweak calculations}}.
  \href{http://dx.doi.org/10.1007/JHEP11(2021)085}{JHEP {\bf 07} (2018)  185},
  \href{http://arxiv.org/abs/1804.10017}{{\tt arXiv:1804.10017 [hep-ph]}}.
  [Erratum: JHEP 11, 085 (2021)].

\bibitem{Denner:2019zfp}
A.~Denner, S.~Dittmaier, M.~Pellen, and C.~Schwan, {\em {Low-virtuality photon
  transitions $\gamma^*\to f\bar f$ and the photon-to-jet conversion
  function}}. \href{http://dx.doi.org/10.1016/j.physletb.2019.134951}{Phys.
  Lett. B {\bf 798} (2019)  134951},
  \href{http://arxiv.org/abs/1907.02366}{{\tt arXiv:1907.02366 [hep-ph]}}.

\bibitem{Denner:1999gp}
A.~Denner, S.~Dittmaier, M.~Roth, and D.~Wackeroth, {\em {Predictions for all
  processes e+ e- ---\ensuremath{>} 4 fermions + gamma}}.
  \href{http://dx.doi.org/10.1016/S0550-3213(99)00437-X}{Nucl. Phys. B {\bf
  560} (1999)  33--65}, \href{http://arxiv.org/abs/hep-ph/9904472}{{\tt
  arXiv:hep-ph/9904472}}.

\bibitem{Denner:2005fg}
A.~Denner, S.~Dittmaier, M.~Roth, and L.~H. Wieders, {\em {Electroweak
  corrections to charged-current e+ e- ---\ensuremath{>} 4 fermion processes:
  Technical details and further results}}.
  \href{http://dx.doi.org/10.1016/j.nuclphysb.2011.09.001}{Nucl. Phys. B {\bf
  724} (2005)  247--294}, \href{http://arxiv.org/abs/hep-ph/0505042}{{\tt
  arXiv:hep-ph/0505042}}. [Erratum: Nucl.Phys.B 854, 504--507 (2012)].

\bibitem{Denner:2006ic}
A.~Denner and S.~Dittmaier, {\em {The Complex-mass scheme for perturbative
  calculations with unstable particles}}.
  \href{http://dx.doi.org/10.1016/j.nuclphysbps.2006.09.025}{Nucl. Phys. B
  Proc. Suppl. {\bf 160} (2006)  22--26},
  \href{http://arxiv.org/abs/hep-ph/0605312}{{\tt arXiv:hep-ph/0605312}}.

\bibitem{Mirkes:1992hu}
E.~Mirkes, {\em {Angular decay distribution of leptons from W bosons at NLO in
  hadronic collisions}}.
  \href{http://dx.doi.org/10.1016/0550-3213(92)90046-E}{Nucl. Phys. B {\bf 387}
  (1992)  3--85}.

\bibitem{Collins:1977iv}
J.~C. Collins and D.~E. Soper, {\em {Angular Distribution of Dileptons in
  High-Energy Hadron Collisions}}.
  \href{http://dx.doi.org/10.1103/PhysRevD.16.2219}{Phys. Rev. D {\bf 16}
  (1977)  2219}.

\bibitem{Ebert:2020dfc}
M.~A. Ebert, J.~K.~L. Michel, I.~W. Stewart, and F.~J. Tackmann, {\em
  {Drell-Yan $q_{T}$ resummation of fiducial power corrections at N$^{3}$LL}}.
  \href{http://dx.doi.org/10.1007/JHEP04(2021)102}{JHEP {\bf 04} (2021)  102},
  \href{http://arxiv.org/abs/2006.11382}{{\tt arXiv:2006.11382 [hep-ph]}}.

\bibitem{Manohar:2017eqh}
A.~V. Manohar, P.~Nason, G.~P. Salam, and G.~Zanderighi, {\em {The Photon
  Content of the Proton}}.
  \href{http://dx.doi.org/10.1007/JHEP12(2017)046}{JHEP {\bf 12} (2017)  046},
  \href{http://arxiv.org/abs/1708.01256}{{\tt arXiv:1708.01256 [hep-ph]}}.

\bibitem{Buckley:2014ana}
A.~Buckley, J.~Ferrando, S.~Lloyd, K.~Nordstr\"om, B.~Page, M.~R\"ufenacht,
  M.~Sch\"onherr, and G.~Watt, {\em {LHAPDF6: parton density access in the LHC
  precision era}}. \href{http://dx.doi.org/10.1140/epjc/s10052-015-3318-8}{Eur.
  Phys. J. C {\bf 75} (2015)  132}, \href{http://arxiv.org/abs/1412.7420}{{\tt
  arXiv:1412.7420 [hep-ph]}}.

\bibitem{Czakon:2010td}
M.~Czakon, {\em {A novel subtraction scheme for double-real radiation at
  NNLO}}. \href{http://dx.doi.org/10.1016/j.physletb.2010.08.036}{Phys. Lett. B
  {\bf 693} (2010)  259--268}, \href{http://arxiv.org/abs/1005.0274}{{\tt
  arXiv:1005.0274 [hep-ph]}}.

\bibitem{Czakon:2011ve}
M.~Czakon, {\em {Double-real radiation in hadronic top quark pair production as
  a proof of a certain concept}}.
  \href{http://dx.doi.org/10.1016/j.nuclphysb.2011.03.020}{Nucl. Phys. B {\bf
  849} (2011)  250--295}, \href{http://arxiv.org/abs/1101.0642}{{\tt
  arXiv:1101.0642 [hep-ph]}}.

\bibitem{Czakon:2014oma}
M.~Czakon and D.~Heymes, {\em {Four-dimensional formulation of the
  sector-improved residue subtraction scheme}}.
  \href{http://dx.doi.org/10.1016/j.nuclphysb.2014.11.006}{Nucl. Phys. B {\bf
  890} (2014)  152--227}, \href{http://arxiv.org/abs/1408.2500}{{\tt
  arXiv:1408.2500 [hep-ph]}}.

\bibitem{Czakon:2019tmo}
M.~Czakon, A.~van Hameren, A.~Mitov, and R.~Poncelet, {\em {Single-jet
  inclusive rates with exact color at $ \mathcal{O} $ ($ {\alpha}_s^4 $)}}.
  \href{http://dx.doi.org/10.1007/JHEP10(2019)262}{JHEP {\bf 10} (2019)  262},
  \href{http://arxiv.org/abs/1907.12911}{{\tt arXiv:1907.12911 [hep-ph]}}.

\bibitem{Bury:2015dla}
M.~Bury and A.~van Hameren, {\em {Numerical evaluation of multi-gluon
  amplitudes for High Energy Factorization}}.
  \href{http://dx.doi.org/10.1016/j.cpc.2015.06.023}{Comput. Phys. Commun. {\bf
  196} (2015)  592--598}, \href{http://arxiv.org/abs/1503.08612}{{\tt
  arXiv:1503.08612 [hep-ph]}}.

\bibitem{Buccioni:2019sur}
F.~Buccioni, J.-N. Lang, J.~M. Lindert, P.~Maierh\"ofer, S.~Pozzorini,
  H.~Zhang, and M.~F. Zoller, {\em {OpenLoops 2}}.
  \href{http://dx.doi.org/10.1140/epjc/s10052-019-7306-2}{Eur. Phys. J. C {\bf
  79} (2019) no.~10, 866}, \href{http://arxiv.org/abs/1907.13071}{{\tt
  arXiv:1907.13071 [hep-ph]}}.

\bibitem{Gehrmann:2011ab}
T.~Gehrmann and L.~Tancredi, {\em {Two-loop QCD helicity amplitudes for $q\bar
  q \to W^\pm \gamma$ and $q\bar q \to Z^0 \gamma$}}.
  \href{http://dx.doi.org/10.1007/JHEP02(2012)004}{JHEP {\bf 02} (2012)  004},
  \href{http://arxiv.org/abs/1112.1531}{{\tt arXiv:1112.1531 [hep-ph]}}.

\bibitem{Bauer:2000cp}
C.~W. Bauer, A.~Frink, and R.~Kreckel, {\em {Introduction to the GiNaC
  framework for symbolic computation within the C++ programming language}}.
  \href{http://dx.doi.org/10.1006/jsco.2001.0494}{J. Symb. Comput. {\bf 33}
  (2002)  1--12}, \href{http://arxiv.org/abs/cs/0004015}{{\tt
  arXiv:cs/0004015}}.

\bibitem{Vollinga:2004sn}
J.~Vollinga and S.~Weinzierl, {\em {Numerical evaluation of multiple
  polylogarithms}}. \href{http://dx.doi.org/10.1016/j.cpc.2004.12.009}{Comput.
  Phys. Commun. {\bf 167} (2005)  177},
  \href{http://arxiv.org/abs/hep-ph/0410259}{{\tt arXiv:hep-ph/0410259}}.

\bibitem{Alwall:2014hca}
J.~Alwall, R.~Frederix, S.~Frixione, V.~Hirschi, F.~Maltoni, O.~Mattelaer,
  H.~S. Shao, T.~Stelzer, P.~Torrielli, and M.~Zaro, {\em {The automated
  computation of tree-level and next-to-leading order differential cross
  sections, and their matching to parton shower simulations}}.
  \href{http://dx.doi.org/10.1007/JHEP07(2014)079}{JHEP {\bf 07} (2014)  079},
  \href{http://arxiv.org/abs/1405.0301}{{\tt arXiv:1405.0301 [hep-ph]}}.

\bibitem{Rubin:2010xp}
M.~Rubin, G.~P. Salam, and S.~Sapeta, {\em {Giant QCD K-factors beyond NLO}}.
  \href{http://dx.doi.org/10.1007/JHEP09(2010)084}{JHEP {\bf 09} (2010)  084},
  \href{http://arxiv.org/abs/1006.2144}{{\tt arXiv:1006.2144 [hep-ph]}}.

\bibitem{Denner:2019vbn}
A.~Denner and S.~Dittmaier, {\em {Electroweak Radiative Corrections for
  Collider Physics}}.
  \href{http://dx.doi.org/10.1016/j.physrep.2020.04.001}{Phys. Rept. {\bf 864}
  (2020)  1--163}, \href{http://arxiv.org/abs/1912.06823}{{\tt arXiv:1912.06823
  [hep-ph]}}.

\end{thebibliography}\endgroup

\end{document}